\documentclass[
  journal=largetwo,
  manuscript=article-type,
  year=2025
]{cup-journal}

\usepackage{amsmath}
\usepackage[nopatch]{microtype}
\usepackage{booktabs}
\usepackage{subfigure}
\usepackage{subcaption}
\usepackage{graphicx}
\usepackage{amssymb}
\usepackage{multirow}

\PassOptionsToPackage{hyphens}{url}\usepackage{hyperref}

\definecolor{dodgerblue}{RGB}{30, 144, 255}

\hypersetup{colorlinks,citecolor=dodgerblue,linkcolor=blue,urlcolor=blue}

\usepackage{xurl}  

\title{Predicting Sunyaev-Zel’dovich effect observations of galaxy cluster cavities with the Square Kilometre Array}

\author{S. Geris}
\affiliation{School of Chemical and Physical Sciences, Victoria University of Wellington, PO Box 600, Wellington 6140, New Zealand}
\email[S. Geris]{sophiageris@gmail.com}

\author{Y. C. Perrott}
\affiliation{School of Chemical and Physical Sciences, Victoria University of Wellington, PO Box 600, Wellington 6140, New Zealand}

\begin{document}

\begin{abstract}
Galaxy cluster X-ray cavities are inflated by relativistic jets that are ejected into the intracluster medium by active galactic nuclei (AGN). AGN jets prevent predicted cooling flow establishment at the cluster centre, and while this process is not well understood in existing studies, simulations have shown that the heating mechanism will depend on the type of gas that fills the cavities. Thermal and non-thermal distributions of electrons will produce different cavity Sunyaev Zel’dovich (SZ) effect signals, quantified by the `suppression factor' $f$. This paper explores potential enhancements to prior constraints on the cavity gas type by simulating suppression factor observations with the Square Kilometre Array (SKA). Cluster cavities across different redshifts are observed to predict the optimum way of measuring $f$ in future observations. We find that the SKA can constrain the suppression factor in the cavities of cluster MS 0735.6+7421 (MS0735) in as little as 4 hours, with a smallest observable value of $f \approx 0.42$. Additionally, while the SKA may distinguish between possible thermal or non-thermal suppression factor values within the cavities of MS0735 if it observes for more than 8 hours, determining the gas type of other clusters will likely require observations at multiple frequencies. The effect of cavity line of sight (LOS) position is also studied, and degeneracies between LOS position and the measured value of $f$ are found. Finally, we find that for small cavities (radius < $80$kpc) at high redshift ($z \approx 1.5$), the proposed high frequencies of the SKA (23.75GHz - 37.5GHz) will be optimal, and that including MeerKAT antennas will improve all observations of this type.
\end{abstract}

\section{Introduction}\label{sec:intro}

Galaxy clusters are the largest known gravitationally collapsed objects in the Universe and are the result of the gravitational evolution of primordial regions of over-density. Studying their astrophysical processes is paramount to understanding the evolution of these structures. The interaction of the intracluster medium (ICM) with the jets of active galactic nuclei (AGN) is one such process. AGN jets are believed to counteract cooling flows, as reviewed by \citet{Bourne}. Extensive studies have been conducted to understand how these jets transfer energy to the ICM, after X-ray spectroscopic observations in cluster Abell 1835 revealed an absence of cooling below 3 keV \citep{peterson2}. While it is likely that a self-regulated feedback loop is the source of heating \citep{McNul}, the energy transfer mechanism remains poorly understood. One process that depicts the considerable impact of the AGN jets on the surrounding ICM is the inflation of large X-ray deficient cavities (or bubbles) by the jets as they course through the plasma. The X-ray emitting thermal ICM is displaced, and the cavities are associated with radio `lobes’ which represent the termination of the jets into the outer regions of the ICM. The X-ray cavities and jets are expected to communicate information about the energetics of the AGN and hence the way the ICM is heated. Possible heating mechanisms include “effervescent” or bubble heating \citep{Abdulla, effervescent}; shock heating \citep{Bourne}; and mixing of the gas in the bubbles with the surrounding ICM \citep{Yang}.

During the development of the theories of heating distributions and AGN feedback cycles discussed above, two outstanding questions have come to light:

\begin{enumerate}
\item What is the composition of the AGN jets that inflate the X-ray cavities?
\item What can the jet composition tell us about the dominant heating processes that are able to counteract the cooling flows?
\end{enumerate}

Broadly, there are two main categories for the jet composition, and in turn, the composition of the cavities they inflate: thermal and non-thermal gas. Momenta/energy of electrons in a thermal gas are described by a Maxwell-Boltzmann distribution, whereas a non-thermal gas typically follows a power law \citep{Enslin}. An environment of thermal gas is possible via the entrainment of local ICM into the jet \citep{Croston, Hardcastle2003, Hardcastle2007} and cosmic rays constitute a non-thermal gas \citep{CRs}. 

Observations have started to constrain the jet and therefore cavity compositions. For example, \cite{Abdullaetal} note that synchrotron emission detected from the cavities suggests that they might consist of magnetic fields and non-thermal relativistic electrons. However, there are some limitations to this description of cavity composition, as the support does not appear sufficient to overcome the weight of the surrounding ICM. Heavy particles (protons and ions) — either entrained from the local ICM or provided by cosmic rays — likely exist within the cavities, to help support them against the overlying gas. \cite{Dunn} and \cite{DeYoung} attempt to explain the origin of these heavy particles, suggesting that cavity plasma is thermal, and entrained from the ICM. In contrast, \cite{Guo} focus on the shape of the cavities, which implies non-thermal cosmic rays are the source of the heavy particles. 

Attempts have also been made to show that thermal and non-thermal cavity contents will imply different dominant heating mechanisms. \cite{Yang} conduct an investigation using three-dimensional hydrodynamic simulations of cluster evolution, including the injection of thermal particles by the jets into the cavities. They find that bubble mixing appears to be a dominant heating source. \cite{Brighenti} simulate the AGN feedback process, by assuming that the jets are composed of non-thermal cosmic rays. In contrast to \cite{Yang}, they find that the heating process must occur independently of the cavities via a ‘postcavity mass outflow’. These studies emphasize the importance of determining the cavity gas type to understanding the cluster heating mechanism.

The Sunyaev-Zel'dovich (SZ) effect has been used by authors \cite{Abdullaetal} and \cite{Orlowski} to observe the cavities of cluster MS 0735.6+7421 (hereafter MS0735) at a redshift of $z = 0.216$, which hosts the largest (radius $\sim$ 100kpc) and most energetic pair of cavities known. Relativistic thermal and non-thermal electron distributions produce distinguishable SZ signals from the surrounding (lower temperature, thermal) ICM, thus signal from the cavities can be described by a `suppression factor' which depends on the cavity gas type.  Attempts to measure this suppression factor in cluster MS0735 were first made by \cite{Abdullaetal} using the Combined Array for Research in Millimeter-wave Astronomy (CARMA) at 30 GHz. They found that there is almost complete suppression of the SZ signal within the cavities, constraining the gas to either be non-thermal, or thermal with electron temperatures of at least several hundred keV. \cite{Orlowski} also attempt to measure the suppression factor in cluster MS0735 with MUSTANG-2 at 90GHz. They find a range of possible suppression factors that are mostly significantly lower than those found in \cite{Abdullaetal}. However, their general findings are in line with the thermal constraint from \cite{Abdullaetal}; that if the bubbles in MS0735 are supported by thermal pressure, the electron temperature must be more than a few hundred keV.

While these studies have significantly advanced the understanding of cavity composition, they both suggest that further exploration of suppression factor measurements is necessary to more accurately determine the behaviour of cavity gas. This includes the use of different instruments to provide higher sensitivity and different frequency ranges to exploit the spectral features of the SZ effect. This paper builds on these existing suppression factor observations by simulating observations of galaxy cluster cavities by the Square Kilometre Array (SKA) \citep{ska1, MSU-CSE-06-2}. The SKA will be the largest radio telescope array in the world and is the largest international project in radio astronomy. It will provide immense improvement to angular resolution, sensitivity and $uv$-coverage, and the combination of long and short baselines are suited to detect small and large scale structures respectively, making it a promising tool to observe cluster bubbles and the extended SZ emission. It therefore has the potential to aid in understanding the jet-ICM physics.

The remainder of this paper is laid out as follows. Section \ref{sec:2} describes the method for simulating cavity observations with the SKA. Section \ref{sec:ms0735} presents an investigation into SKA simulated observations of cluster MS0735 including an exploration of the effect of cavity line of sight position on the suppression factor measurement, while section \ref{sec:freq} explores multiple SKA frequency observations of a more diverse range of clusters. Improvements to the results collected are explored in section \ref{sec:improv} by including the MeerKAT antennas in the SKA telescope array. Finally, section \ref{sec:future} discusses further research to be done in this area and our conclusions are summarised in section \ref{sec:end}. A $\Lambda$CDM cosmology is assumed throughout this work, with $\Omega_{\Lambda}=0.7$, $\Omega_\mathrm{m}=0.3$, and $H_\mathrm{0}=70$ km s$^{-1}$ Mpc$^{-1}$. 

\section{Simulating SKA Observations}\label{sec:2}

Galaxy clusters are modelled by calculating the SZ signal in each pixel that corresponds to a small area patch of the cluster’s projection on a simulated 2D sky. The process is described below.

\subsection{Global ICM model}\label{sec:icm}
\label{} 

The change in intensity of the cosmic microwave background (CMB) photons as they are scattered by the ICM electrons is given by

\begin{equation}
\Delta I_{\nu} \approx i_{0}y\frac{x^{4}e^{x}}{(e^{x}-1)^{2}}\left(x \frac{e^{x}+1}{e^{x}-1}-4\right) \equiv i_{0}yg(x),
\label{eq:I_nu}
\end{equation}

where $x = \frac{h \nu}{kT_\mathrm{CMB}}$, $g(x)$ is the frequency dependence, $y$ is the Compton-$y$ parameter and $i_{0} = \frac{2(k T_\mathrm{CMB})^{3}}{(hc)^{2}}$ is the CMB intensity. We assume the standard non-relativistic approximation to the thermal SZ effect signal for $g(x)$, given by

\begin{equation}
g(x) = \frac{x^{4}e^{x}}{(e^{x}-1)^{2}}\left(x \frac{e^{x}+1}{e^{x}-1}-4\right).
\label{eq:gx}
\end{equation}

The Compton-$y$ parameter is given by 

\begin{equation}
y(\theta_\mathrm{proj}) = \frac{\sigma_\mathrm{T}}{m_\mathrm{e}c^{2}} \int^{\infty}_{-\infty} P_\mathrm{e}\left(\sqrt{\theta_\mathrm{proj}^{2} + l^{2}}\right) \, \mathrm{d}l,
\label{eq:ytheta}
\end{equation}

where $\theta_\mathrm{proj} = \sqrt{\mathrm{x}^{2} + \mathrm{y}^{2}}$ is a projected radius with $\mathrm{x}$ and $\mathrm{y}$ in arcmin representing coordinates on a 2D pixel grid of the sky projection. The parameter $\sigma_\mathrm{T}$ is the Thomson cross section, $m_\mathrm{e}$ is the electron mass and $c$ is the speed of light. The projected radius $\theta_\mathrm{proj}$ describes the radius of each pixel in the grid from the centre of the cluster (defined at $(\mathrm{x}, \mathrm{y})$ $= (0,0)$). We assume that $P_\mathrm{e}$, the thermal pressure in the cluster, is given by the commonly-used Generalised Navarro-Frenk-White (GNFW) pressure profile \citep{Nagai}

\begin{equation}
P_\mathrm{e}(r) = \frac{P_\mathrm{ei}}{(r/r_\mathrm{s})^{\gamma}[1+(r/r_\mathrm{s})^{\alpha}]^{(\beta-\gamma)/ \alpha}},
\label{eq:gnfw}
\end{equation} 

where $P_\mathrm{ei}$ is the normalisation coefficient, $r_{s}$ is a characteristic radius and $\alpha$, $\beta$, $\gamma$ are shape parameters that describe the slopes of the pressure profiles. We assume the universal shape parameter values given in \cite{Arnaud}. We obtain $P_\mathrm{ei}$ by considering the total integrated Comptonisation parameter, $Y_\mathrm{tot}$ given by integrating 

\begin{equation}
Y(r) = \frac{\sigma_\mathrm{T}}{D_\mathrm{A}^{2}m_\mathrm{e}c^{2}} \int^{r}_\mathrm{0} P_{e}(r) 4 \pi r^{2} \, \mathrm{d}r,
\label{eq:ytotint}
\end{equation}

out to $r=\infty$. In the GNFW model, $Y_\mathrm{tot}$ has an analytical solution

\begin{equation}
Y_\mathrm{tot} = \frac{4 \pi \sigma_\mathrm{T}}{m_\mathrm{e}c^{2}} P_\mathrm{ei} D_\mathrm{A} \theta_\mathrm{s}^{3} \frac{\Gamma(3-\gamma) \Gamma(\frac{\beta -3}{\alpha})}{\alpha \Gamma(\frac{\beta - \gamma}{\alpha})},
\label{eq:ytot}
\end{equation}

which can be rearranged for $P_\mathrm{ei}$, given $Y_\mathrm{tot}$. In this equation, $D_\mathrm{A}$ is the angular diameter distance to the cluster and $\theta_\mathrm{s}$ = $r_\mathrm{s}/D_\mathrm{A}$ is the angular equivalent of a physical characteristic radius $r_\mathrm{s} = r_\mathrm{500}/c_\mathrm{500}$. The radius $r_\mathrm{500}$ surrounds a volume of the cluster which has a mean density 500 times the critical density of the universe at that redshift, and is the physical radius of interest. The gas concentration parameter, $c_\mathrm{500}$, links to a physical description of a cluster with mass \citep{Arnaud}, and $\Gamma$ is the gamma function. Given $Y_\mathrm{500}$, it is possible to calculate $Y_\mathrm{tot}$ by taking equation \ref{eq:ytotint}, integrated out to $\infty$, and dividing by the same integral, integrated out to $r_{500}$. This is then multiplied by $Y_\mathrm{500}$.

A scaling relation is calibrated between $Y_\mathrm{500}$ and $M_\mathrm{500}$ in \cite{scalingrelation},

\begin{equation}
E^{-2/3}(z) \left[\frac{D_\mathrm{A}^{2}Y_\mathrm{500}}{10^{-4}Mpc^{2}}\right] = 10^{-0.175 \pm 0.011}\left[\frac{M_\mathrm{500}}{6 \times 10^{14}M_\odot}\right]^{1.77 \pm 0.06},
\label{eq:scalingrel}
\end{equation}

where $M_\mathrm{500}$ is the mass within $r_\mathrm{500}$ and $E(z)$ is the cosmological function used to calculate the Hubble parameter $H(z)$. This relation is used to find $Y_\mathrm{500}$, given $M_\mathrm{500}$ and $z$. 

We assume the longer baselines of the SKA, insensitive to the large-scale SZ emission, will be used to accurately subtract compact sources so that cluster and cavity detections are not impacted.

\subsection{Cavity model}\label{sec:cavity}
The next step in the model construction is to embed the cavities into the global ICM. In all of the simulated observations in this paper, we consider two cavities. Since we are considering a binary description of cavity gas in terms of (relativistic) thermal or non-thermal distributions, we want to depict how the imprints on the scattered photon spectrum will differ, so that their SZ signals can be distinguished from the global ICM. A general formulation of the SZ effect is described in detail in \cite{Enslin} and \cite{Colafrancescoetal} which can be adapted to suit the particular scattering electrons' distribution. A brief description is provided here. 

The change in flux density due to the scattering of CMB photons by electrons within the cavities in the optically thin limit is given by,

\begin{equation}
\delta i(x) = (j(x)-i(x))\tau_\mathrm{cav},
\label{eq:deltai}
\end{equation}

where $i(x)$ is the shape of the undistorted CMB specific intensity, $i(x)\tau_\mathrm{cav}$ is the flux scattered to other frequencies from $x$ and $j(x)\tau_\mathrm{cav}$ is the flux scattered from other frequencies to $x$. Note that this is for single scattering only. The optical depth, $\tau_\mathrm{cav}$, is given by,

\begin{equation}
\tau_\mathrm{cav} = \sigma_\mathrm{T} \int_\mathrm{cav} n_\mathrm{e}\, \mathrm{d}l,
\end{equation}

where the integration limits $\mathrm{cav}$ (notation from \citealt{Abdullaetal}) represents the physical boundary that separates the cavity from the global ICM. Equation \ref{eq:deltai} can also be written as,

\begin{equation}
\delta i(x) = \tilde{g}(x)y_\mathrm{cav},
\label{eq:gytilde}
\end{equation}

where,

\begin{equation}
y_\mathrm{cav} = \frac{\sigma_\mathrm{T}}{m_\mathrm{e}c^{2}}\int_\mathrm{cav} P_\mathrm{e} \, \mathrm{d}l,
\label{eq:ytilde}
\end{equation}

\begin{equation}
\tilde{g}(x) = (j(x) -i(x)) \frac{m_\mathrm{e}c^{2}}{\langle k \tilde{T}_\mathrm{e}\rangle},
\label{eq:gtildex}
\end{equation}

\begin{equation}
k \tilde{T}_\mathrm{e} \equiv \frac{P_\mathrm{e}}{n_\mathrm{e}},
\label{eq:k}
\end{equation}

\begin{equation}
\langle k \tilde{T}_\mathrm{e}\rangle = \frac{\int n_\mathrm{e} k \tilde{T}_\mathrm{e}\, \mathrm{d}l}{\int  n_\mathrm{e} \, \mathrm{d}l}.
\end{equation}

\begin{figure}
\centering
\includegraphics[width=\columnwidth]{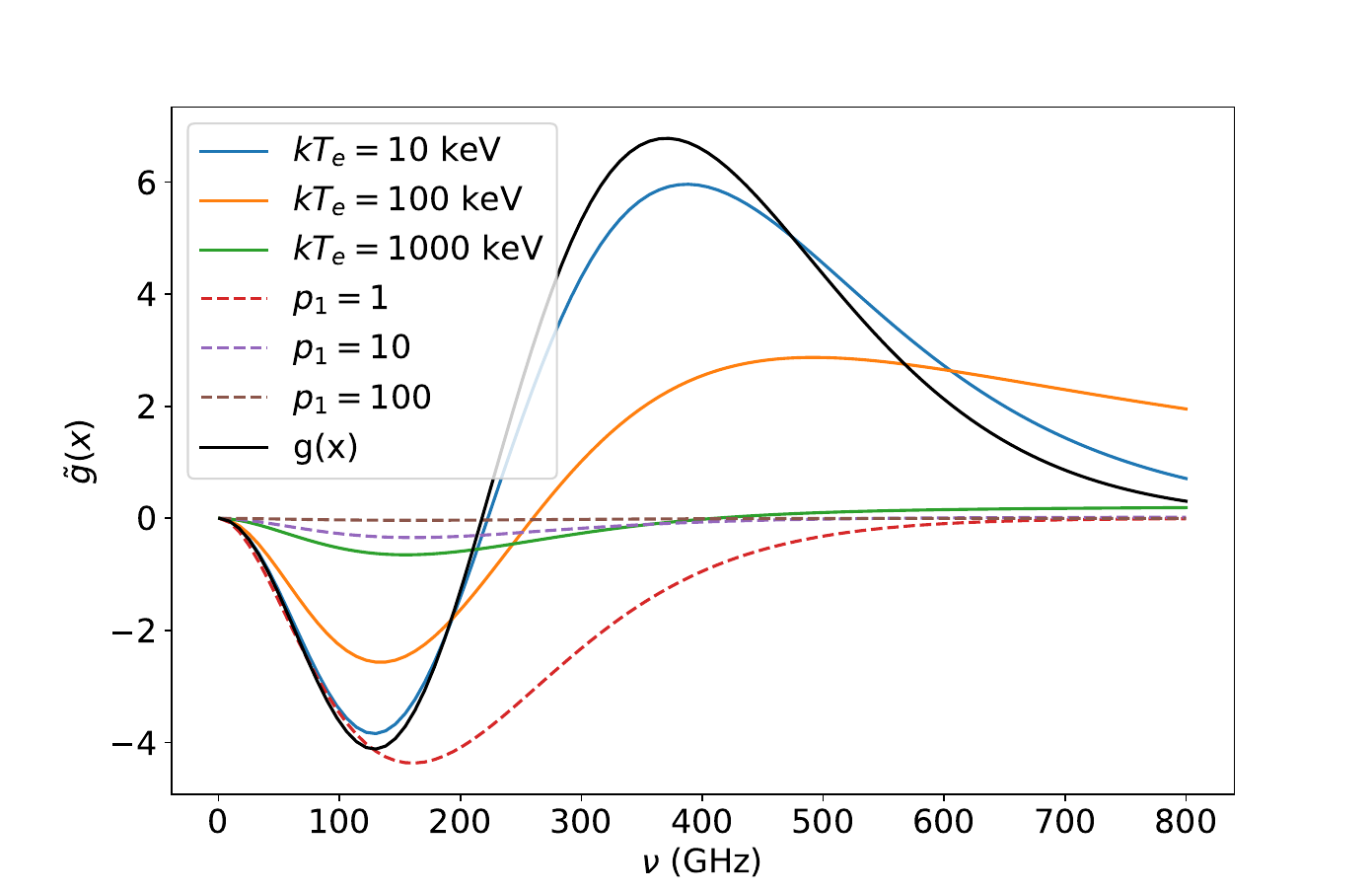}
\caption{Dashed lines depict the distorted spectrum of SZ signal from a non-thermal distribution of cosmic ray electrons with minimum momenta $p_\mathrm{1} = 1$, $10$, or $100$, maximum momentum $p_\mathrm{2}=10^{5}$, and $\alpha=6$. The solid lines depict distortions from thermal distributions of ICM entrained free electrons with temperature $k T_\mathrm{e} = 10$, $100$ or $1000$ keV. The solid black curve is $g(x$) of the global ICM thermal SZ signal. The different distortions depicted in the curves with different values of $kT_\mathrm{e}$ and $p_\mathrm{1}$ mean it is possible to distinguish between a thermal and non-thermal signal at different frequencies.}
\label{fig:gtilde}
\end{figure}

Equation \ref{eq:gytilde} is analogous to the non-relativistic thermal case of the global ICM in equation \ref{eq:I_nu}. Equation \ref{eq:ytilde} defines the Compton-$y$ parameter, where the segment of the line of sight that is being integrated over must penetrate a bubble. The distorted SZ spectrum, $\tilde{g}(x)$, describes the signal received from the bubbles for different frequencies. It is dependent on the type of gas which is doing the scattering. Some examples of $\tilde{g}(x)$ are shown in Figure \ref{fig:gtilde}. The thermodynamic temperature for the case of a thermal electron population, $k \tilde{T}_\mathrm{e}$, is given by equation \ref{eq:k}). The concept of temperature is more ambiguous in a non-thermal gas, therefore $k \tilde{T}_\mathrm{e}$ is the pseudo-temperature of the particles. For a population of non-thermal cosmic ray electrons described by a power-law distribution, $P_\mathrm{e}$  in equation \ref{eq:k} is specific to the cosmic ray electron pressure, and is given by,

\begin{equation}
P_\mathrm{CR_\mathrm{e}} = \frac{n_\mathrm{CR_\mathrm{e}} m_\mathrm{e} c^{2} \left(\alpha -1\right)}{6\left[p^{1-\alpha}\right]_{p_\mathrm{2}}^{p_\mathrm{1}}} \left[B_\mathrm{\frac{1}{1+p^{2}}} \left(\frac{\alpha -2}{2}, \frac{3- \alpha}{2}\right)\right]^{p_\mathrm{1}}_{p_\mathrm{2}},
\label{eq:pcre}
\end{equation}

\citep{Pfrommer}. This is used to estimate the constant value of $k \tilde{T}_{e}$ for use in equation \ref{eq:gtildex}. In equation \ref{eq:pcre}, $\alpha$ is the power-law index of the momentum distribution, $B$ is the incomplete Beta function, $n_{CR_{e}}$ is the cosmic ray electron number density, and $p=\beta_{e}\gamma_{e}$ is the normalised electron momentum \citep{Enslin}. The quantity $\beta_{e}$ is electron velocity relative to the speed of light and $\gamma_{e}$ is the Lorentz factor, therefore $p$ is dimensionless. The parameters $p_\mathrm{1}$ and $p_\mathrm{2}$ are minimum and maximum momenta of the electrons in the distribution. The notation

\begin{equation}
[f(p)]_\mathrm{p_\mathrm{2}}^{p_\mathrm{1}} = f(p_\mathrm{1}) - f(p_\mathrm{2}),
\end{equation}

is used in equation \ref{eq:pcre}. 

In the literature, it is commonly assumed that the bubbles and surrounding ICM are in pressure equilibrium, as simulations have found that after their inflation, the bubbles will expand until they reach pressure equilibrium with their surroundings (e.g. \citealt{equilibrium}). Therefore, we assume that the shape of the pressure profile within the cavities follows the general shape of the global ICM. This means that the Compton-$y$ parameter within the bubbles, described by equation \ref{eq:ytilde}, will be an integral over the GNFW pressure (equation \ref{eq:gnfw}).

The scattered spectrum $j(x)$ in equation \ref{eq:deltai} and \ref{eq:gtildex} is given by, 

\begin{equation}
j(x) = \int_\mathrm{0}^{\infty}  K(t)i(x/t) \, \mathrm{d}t,
\end{equation}

where $K(t)$ is the photon redistribution function, giving the probability that a photon is scattered to a frequency $t$ times its original frequency. The photon redistribution function is dependent on the type of gas that the photons are scattered off and is given by,

\begin{equation}
K(t) = \int^{\infty}_\mathrm{0} f_\mathrm{e}(p)\mathcal{P}(t;p) \, \mathrm{d}p,
\end{equation}

where $f_\mathrm{e}(p)$ is the electron momentum distribution. This is where the differentiation between a thermal and non-thermal distribution comes into the formulation. The photon redistribution function for a mono-energetic electron distribution, $\mathcal{P}(t;p)$, has a compact analytical solution derived in \cite{Enslin}.

If the characterisation of gas within the bubbles is thermal via ICM entrainment, then $f_\mathrm{e}(p)$ is given by a relativistic form of the Maxwell-Boltzmann distribution

\begin{equation}
f_\mathrm{e, th}(p) = \frac{\beta_\mathrm{th}}{K_\mathrm{2}(\beta_\mathrm{th})}p^{2}\exp(-\beta_\mathrm{th}\sqrt{1+p^{2}}),
\label{eq:3}
\end{equation}

where $K_\mathrm{2}$ is the modified Bessel function and $\beta_\mathrm{th} = m_\mathrm{e}c^{2}/kT_\mathrm{e}$. Note that while massive clusters have global ICM temperatures of $\sim10$ keV, and therefore, our global ICM model should technically also follow this form with a mass-dependent temperature, the spectral changes at the frequencies we are considering for the SKA (14.11 - 37.5 GHz) are very small for these lower temperatures compared to the more significant spectral changes for the $\sim100$ keV temperatures that are investigated in the later sections of this paper. This can be seen in Figure \ref{fig:gtilde}, where the blue curve for $\tilde{g}(x)$ at $kT_{e}$ = 10 keV follows the black curve for $g(x)$ very closely at frequencies below $\sim 100$ GHz, compared to the orange curve for $kT_{e}$ which is more distinct.

A non-thermal population of cosmic ray electrons inhabiting the bubbles will be described by a single power-law electron momentum distribution, given by,

\begin{equation}
f_\mathrm{e, non-th}(p; \alpha, p_\mathrm{1}, p_\mathrm{2}) = \frac{(\alpha -1)p^{- \alpha}}{p_\mathrm{1}^{1-\alpha}-p_\mathrm{2}^{1-\alpha}}; p_\mathrm{1}<p<p_\mathrm{2},
\label{eq:4}
\end{equation}

where $\alpha$ is the power-law index. The parameter $p_\mathrm{1}$ is considered a free variable, as the amount of scattering will be primarily determined by its value \citep{Colafrancescoetal}. The effect of changing $p_\mathrm{2}$ is small, so its value is kept fixed at $p_\mathrm{2}=10^{5}$ 
\citep{Colafrancescoetal}. 

Given the assumption of consistent pressure profiles inside and outside cavities, the relative change in flux density from the whole cluster can be redefined as the standard thermal SZ contribution from the cluster with a spherical cavity removed, plus any contribution from the general particle population inside the cavity. This is given by,

\begin{equation}
\delta i(x) = [y_\mathrm{cl} - y_\mathrm{cav}]g(x) + y_\mathrm{cav}\tilde{g}(x),
\label{eq:clcav}
\end{equation}

where the Comptonization from the GNFW model with line of sight through the entire cluster is $y_\mathrm{cl}$. The integration limits of $y_\mathrm{cav}$ (equation \ref{eq:ytilde}) are defined by the intercepts between the line of sight and the cavity boundaries, given by

\begin{equation}
z = \pm\sqrt{r_\mathrm{b}^{2} - (\mathrm{x}-\mathrm{x}_\mathrm{b})^{2} - (\mathrm{y}-\mathrm{y}_\mathrm{b})^{2}} + z_\mathrm{b},
\label{eq:z}
\end{equation}

where $r_\mathrm{b}$ is the radius of the bubble, $\mathrm{x}$ and $\mathrm{y}$ are the coordinates on the pixel grid, $\mathrm{x}_\mathrm{b}$ and $\mathrm{y}_\mathrm{b}$ are the coordinates of the bubble centre on the pixel grid and $z_\mathrm{b}$ is the amount that the bubble is shifted in or out of the plane, along the line of sight.

\begin{figure}
  \begin{subfigure} 
    \centering
    \includegraphics[width=0.8\columnwidth]{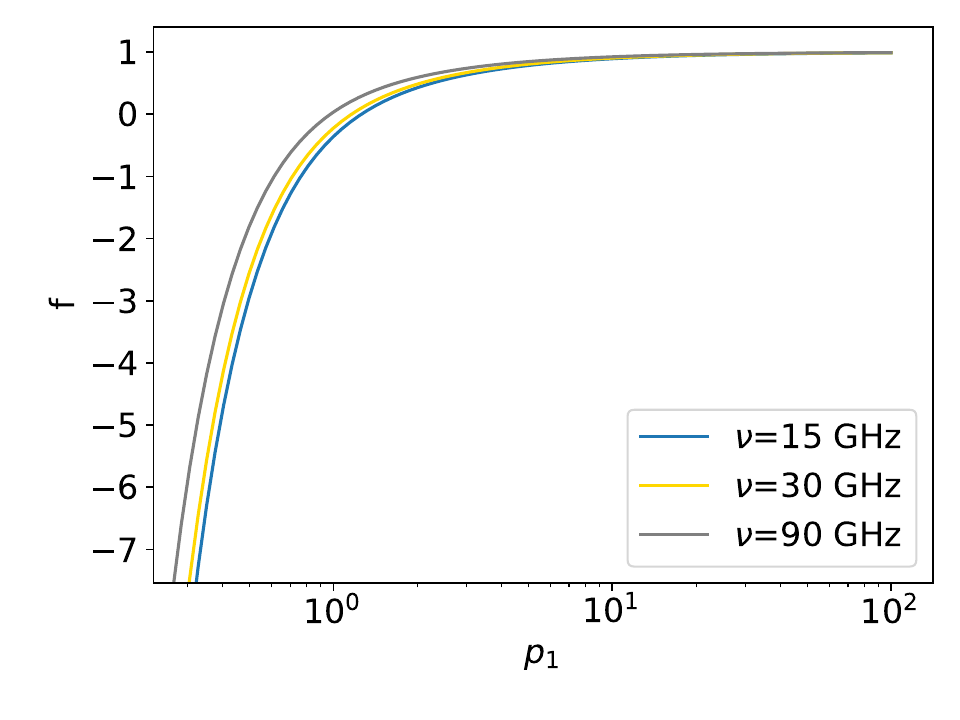}
  \end{subfigure}%
  \\
  \begin{subfigure} 
    \centering
    \includegraphics[width=0.8\columnwidth]{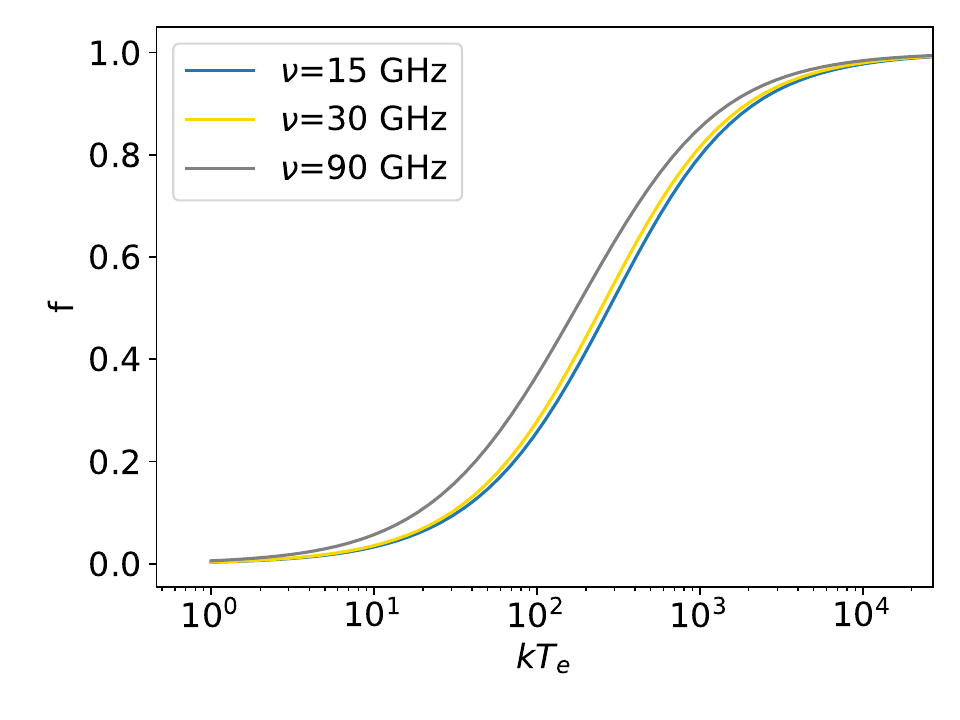}
  \end{subfigure}
  \caption{Top figure: The suppression factor (equation \ref{eq:supp}) of a non-thermal distribution of electrons in the cluster cavities of MS0735 ($\alpha = 6$ and $p_\mathrm{2} = 10^{5}$). Bottom figure: The suppression factor of a thermal distribution of electrons with temperature $kT_\mathrm{e}$ in keV. The frequencies are those of some SZ instruments considered in this paper: SKA at $\sim 15$ GHz, CARMA at $30$ GHZ and GBT MUSTANG-2 at $90$ GHz. The suppression factor of a non-thermal gas can be negative, meaning an increased SZ signal, compared to a thermal gas which must be $\geq 0$. At the frequencies shown here, $f>1$ is not possible.}
  \label{fig:f}
\end{figure}

Equation \ref{eq:clcav} can be simplified to:

\begin{equation}
\delta i(x) = (y_\mathrm{cl} - fy_\mathrm{cav})g(x), 
\label{eq:if}
\end{equation}

where 

\begin{equation}
f \equiv 1 - \frac{\tilde{g}(x)}{g(x)}, 
\label{eq:supp}
\end{equation}

is defined by \cite{Abdullaetal} as the cavity suppression factor. The function $f$ describes a scaled version of the thermal non-relativistic SZ effect in cavity regions of the ICM, due to the different scattering caused by either relativistic thermal or non-thermal electrons within them. Since $f$ depends on $\tilde{g}(x)$ (equation \ref{eq:gtildex}), then it will differ for a thermal or non-thermal cavity gas. The top and bottom images in Figure \ref{fig:f} depict the suppression factor of a non-thermal and thermal gas, respectively, at different frequencies.

The coordinates $z$ in equation \ref{eq:z} are used as the bounds of the integral \ref{eq:ytheta} when $((\mathrm{x}-\mathrm{x}_\mathrm{b})^{2} + (\mathrm{y}-\mathrm{y}_\mathrm{b})^{2}) < r_\mathrm{b}^{2}$, and the integral is multiplied by $f$ (Equation \ref{eq:supp}) to obtain the effective $y$-map of the cluster and cavities,

\begin{equation}
\begin{split}
y = \frac{\sigma_\mathrm{T}}{m_\mathrm{e}c^{2}} \int^{\infty}_{-\infty} P_\mathrm{e}(\sqrt{\theta_\mathrm{proj}^{2} + l^{2}}) \, \mathrm{d}l  - \\ f\left(\frac{\sigma_\mathrm{T}}{m_\mathrm{e}c^{2}} \int^{+z}_{-z} P_\mathrm{e}(\sqrt{\theta_\mathrm{proj}^{2} + l^{2}})\, \mathrm{d}l\right).
\end{split}
\label{eq:final}
\end{equation}

Depending on whether a cluster is being modelled with thermal or non-thermal gas within the bubbles, equations \ref{eq:3} or \ref{eq:4} are used respectively, to calculate $f$. The $y$-map is then converted to a signal map in MJy/sr by multiplying equation \ref{eq:final} by the non-relativistic thermal spectral distortion $g(x)$, and $i_\mathrm{0} = \frac{2(k T_\mathrm{CMB})^{3}}{(hc)^{2}}$. This can subsequently be converted to a map in Jy/pix for use in interferometric simulations.

\subsection{Interferometric Simulations}

To generate simulations of SKA observations, \textsc{Profile} Software \citep{profile} is used. This software directly mimics the observation process of a real interferometer. The model image of the sky in Jy/pix that is simulated via the process in section \ref{sec:icm} and \ref{sec:cavity} is read from a FITS (Flexible Image Transport System) file. Information including the pixel size in degrees, the cluster location in right ascension (RA) and declination, and the pixels corresponding to that location, is stored in the FITS headers before the file is used as input in \textsc{Profile}. This image represents the sky brightness distribution received by the interferometer. The sky brightness distribution is then modified by the primary beam. The primary beam applied to the simulations was calculated based on an edge-tapered Gaussian aperture illumination function (K. Grainge, priv. comm.) which resulted in a primary beam with a Gaussian central lobe with full-width at half-maximum (FWHM) given in Table \ref{tab:skabands}.  When analyzing the simulations, for simplicity a Gaussian model with the given FWHM was assumed. No calibration errors are assumed in the simulations in this work. The Fourier transform of this map is taken, then sampled at the points in the $uv$-plane corresponding to the expected layout of the SKA antennas to simulate the visibilities. Since the SKA will consist of two arrays (SKA-Mid and SKA-Low), we specify that the simulations presented in this paper will assume the SKA-Mid (referred to as SKA for the remainder of this paper) layout. We will also involve the MeerKAT antennas in section \ref{sec:improv}. See table \ref{tab:skainfo} for the technical details of these instruments and table \ref{tab:skabands} for the proposed frequency bands that are used in this work along side the FWHM of the assumed primary beam. While the SKA bands are not ideal for SZ observations, the SZ effect of galaxy clusters does become visible in the higher end of band 5 which is the frequency range that the simulations in this paper will focus on.

\begin{table}[hbt!]
\begin{threeparttable}
\caption{Technical information of the SKA-Mid and MeerKAT arrays from \cite{ska1} and \cite{MSU-CSE-06-2}.}
\label{tab:skainfo}
\begin{tabular}{lll}
\toprule
\headrow & SKA-Mid & MeerKAT\\
\midrule
Frequency range &  350 MHz-15.4 GHz & 0.58-3.05 GHz\\
Antennas & 133 15m dishes & 64 13.5m dishes\\
Maximum baseline & 150 km & 8 km\\
Location & South Africa & South Africa \\
\bottomrule
\end{tabular}
\end{threeparttable}
\end{table}

\begin{table}[hbt!]
\begin{threeparttable}
\caption{Proposed frequency bands of the SKA alongside the primary beam FWHM assumed in the simulated observations in this work. Band 5+ and 6 would come with a possible expansion of SKA1 (the first phase of the SKA project) \citep{band6}.}
\label{tab:skabands}
\begin{tabular}{llll}
\toprule
\headrow SKA-Mid Band & Frequency Range (GHz) & Primary Beam FWHM (arcmin) \\
\midrule
Band 5b & 8.3-15.3 & 5.66\\
Band 5+ & 22.5-25 & 3.37\\
Band 6 & 36.25-38.75 & 2.13\\
\bottomrule
\end{tabular}
\end{threeparttable}
\end{table}

Gaussian noise is added to the visibilities at a level determined using the SKAO (SKA Observatory) sensitivity calculator (\url{https://www.skao.int/en/ska-sensitivity-calculators}). The noise level used for 14.11 GHz is $0.0061$ Jy (for a single baseline). After the simulation has run and noise has been added, the output is a FITS file of sampled visibilities. A $clean$ algorithm is run to deconvolve the estimate of the observed sky brightness distribution from the dirty image using CASA (Common Astronomy Software Applications). Throughout this work, natural weighting is used to image the interferometric data, giving constant weights to all visibilities since the noise is added at a consistent level to all visibilities.

\subsection{Statistical Analysis}

To determine how well the suppression factor is detected in the clusters involved in this investigation, \textsc{McAdam} software \citep{mcadam} is used to perform Bayesian inference for parameter estimation and model comparison using nested sampling via \textsc{MultiNest} \citep{Feroz_2009, Feroz_2008, Feroz_2019}. A single suppression factor across both cavities is assumed. Because our aim is to investigate the SKA’s ability to observe cluster cavities, setting prior distributions that are not informative is important to determine how well the true parameter values can be recovered. This is particularly important to determine if the SKA will be able to measure distinguishable $f$ values for thermal and non-thermal gas. Uniform priors are mostly used, with a wide range from the minimum and maximum values allowed by the models (see Figure \ref{fig:f}). Table \ref{tab:params} lists the priors used.

To quantify the strength of the detection of the bubbles the Bayesian evidence is used. The Jeffrey's scale \citep{Jeffreys} is often used in Astrophysics to quantify detection via the evidence, and it is also laid out in \cite{kass}. The scale ranks values of the Bayes factor, which is the ratio of the evidences of two models $Z_\mathrm{1}/Z_\mathrm{2}$, where a higher value represents a better fit of model 1 to the data, over model 2. The \textsc{McAdam} software outputs $\ln(Z)$. Therefore, the scale in \cite{kass} converted to $\ln(Z_\mathrm{1}/Z_\mathrm{2})$ is given in table \ref{tab:evidence}. In this work, the model associated with $Z_\mathrm{1}$ is a model of a cluster with bubbles, and a cluster with no bubbles is associated with $Z_\mathrm{2}$. 

\begin{table}[hbt!]
\begin{threeparttable}
\caption{The scale used to quantify a good detection via the Bayesian evidence $Z$ (from \citealt{kass}), when performing model comparison.}
\label{tab:evidence}
\begin{tabular}{ll}
\toprule
\headrow $\ln (Z_\mathrm{1}/Z_\mathrm{2})$ & Detection strength\\
\midrule
<1 & No detection\\
1 to 3 & Positive\\
3 to 5 & Strong \\
$>$5 & Very strong \\
\bottomrule
\end{tabular}
\end{threeparttable}
\end{table}

The final stage of the analysis is to determine how well an observation constrains the suppression factor within the cavities. To determine if the posteriors are a genuine representation of the actual constraints on the parameters that one can infer from the data, given our model, we employ a test taken from \cite{Posteriorvalidation}. This test involves obtaining the total probability mass contained in the highest probability density (HPD) region of the posteriors. Given a data point $x$ sampled from a distribution $f(x)$, the total probability mass, $\zeta(x)$, in the HPD is given by, 

\begin{equation}
\zeta(x) = \int_{f(u)\geq f(x)}f(u) \, \mathrm{d^{n}}u,
\end{equation}

where the region of integration is defined by the constraint $f(u)\geq f(x)$, which is the HPD. The HPD has the property that any data point sampled outside of it will always have a lower probability than the samples within. If a multidimensional variable $\mathbf{x}$ is considered, the mapping to the probability mass $\zeta (\mathbf{x})$ is given by $\mathbb{R}^{n} \rightarrow [0,1]$, and is obtained by calculating $\zeta$ contained within the HPD and having $\mathbf{x}$ as its boundary. 

The null hypothesis is that the data $\mathbf{x}$ are sampled from the distribution $f(\mathbf{x})$. Therefore, the proposed validation procedure is then to make use of the fact that under the null hypothesis, the probability mass $\zeta$ is uniformly distributed in the range $[0,1]$. A proof of this can be found in \cite{Posteriorvalidation}. 

\vspace{5mm} This is utilised in this analysis by testing if the true parameters $\theta$ of a simulated cluster can be sampled accurately from the predicted posteriors. The expression to compute the probability mass becomes,

\begin{equation}
\zeta (\theta) = \int_{p(u|\hat{\theta}) \geq p(\theta|\hat{\theta})} p(u|\hat{\theta})\, \mathrm{d^{n}}u,
\end{equation}

\citep{posterior2} for sample $u$, where $\hat{\theta}$ is the observation and $p(\theta|\hat{\theta})$ is the posterior. Then for multiple realisations of a simulated SKA observation that has been analysed via MultiNest, the values of $\zeta(\theta)$ from each realisation should be uniformly distributed if the posteriors are a truthful representation of the actual constraints on the parameters from the data. The empirical cumulative distribution function (ECDF) for $\zeta(\theta)$ is then expected to be a linear plot, which we refer to as the posterior validation curve. The linear posterior validation curve implies that the true parameter value should be within the 99\% confidence interval of 99\% of the posterior realisations, within the 68\% confidence interval of 68\% of the posteriors, etc. If the curve falls below this expected line, then the errors of the posteriors of the different realisations have been underestimated, and the posteriors are not a good representation of the measurements of the data. If the errors have been overestimated, the curve will fall above the expected line, indicating that the parameter constraints fall closer to the true value than expected if the posterior was being dominated by the likelihood.

Throughout the analysis in this work, either 20 or 50 realisations of each simulated SKA observation are performed. For the first model we tested (section \ref{sec:ms0735}), we inspected the posterior validation curves and shapes of the individual posteriors for the 50 realizations each of a range of different observing times to determine the required observation length for an accurate and informative constraint on the suppression factor.  We found that an average (over the 50 realizations) log-evidence difference $\ln(Z_1/Z_2) \gtrsim 10$ and average suppression factor posterior width $\sigma \lesssim 0.3$ for a given observing time satisfied these requirements.  These metrics were therefore used to define the required observation lengths for subsequent models, with the exception of section 3.3 where $\sigma \lesssim 0.2$ was used instead given the lower suppression factor in the model.

\section{Observing Cluster MS 0735.6+ 7421}\label{sec:ms0735}

\subsection{Modelling MS0735 with CARMA Constraints}
Galaxy cluster MS 0735.6+ 7421 (MS0735), located at $z=0.216$ and with mass $M_\mathrm{500}=8 \times 10^{14}M_{\odot}$ \citep{Gitti07}, is chosen as the first candidate for simulating X-ray cavity observations with the SKA. This cluster hosts the largest and most energetic X-ray cavities currently known, making it an excellent observation target. The redshift of MS0735 corresponds to a scale of 3.503 kpc/arcsec and we observe at $14.11$ GHz.

Previous observations of the SZ signal from cluster MS0735 by CARMA and MUSTANG-2 have been used to constrain some of the cavity parameters for the model image of the sky. Their measured suppression factors are used to determine the corresponding temperature $kT_\mathrm{e}$ or minimum momentum $p_\mathrm{1}$ within the cavities used in our models. We begin with the CARMA constraints. Adopting the notation $f_{\nu}$ where $\nu$ is the observation frequency, CARMA measured $f_{30}=0.98$ (assuming the cavity minor axis in the elliptical model from \cite{Abdullaetal} falls along the line of sight) which corresponds to $p_\mathrm{1} \approx 100$ in the case of a non-thermal gas, and $kT_\mathrm{e} \approx 10000$ keV for a thermal gas. These produce $f_{14} = 0.977$ and $f_{14} = 0.989$ for thermal and non-thermal gas, respectively. Because these suppression factors are so similar, the analysis going forward is for non-thermal bubbles with suppression factor $f_{14}=0.989$. Later in this section we discuss the SKA's potential to distinguish between a thermal and non-thermal bubble scenario. The other parameters that are used in the model image of cluster MS0735 are listed in table \ref{tab:params}. Alongside these are the prior types and values that are used in the Bayesian analysis. Note that non-physical values are included in the prior range for the suppression factor because the true suppression factor is very close to 1 ($f_{14} = 0.989$), therefore it is very likely that the measured value would fall in the non-physical prior range due to noise from the observation.

\begin{figure}
\centering
\includegraphics[width=0.9\columnwidth]{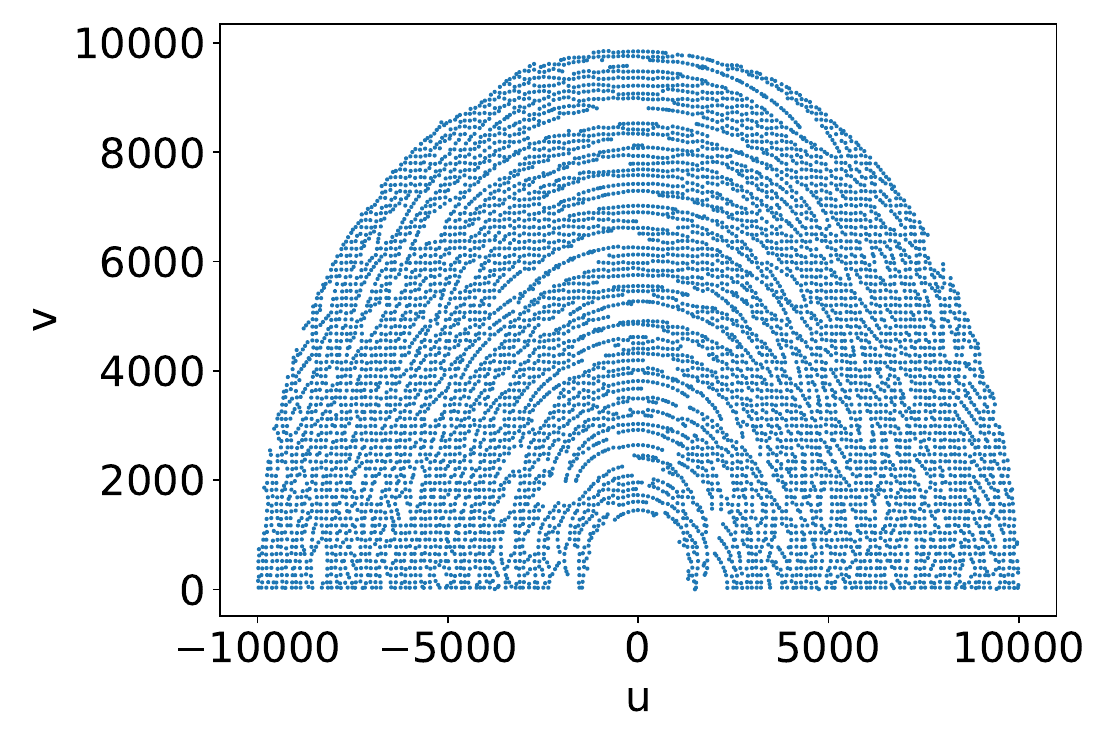}
\caption{The uv coverage in units of $\lambda$ (with a cut-off of 10000$\lambda$) of a simulated 8 hour SKA observation.}
\label{fig:uvv}
\end{figure}

\begin{figure*}
		\begin{subfigure}
                \centering
        	\includegraphics[width=0.8\linewidth]{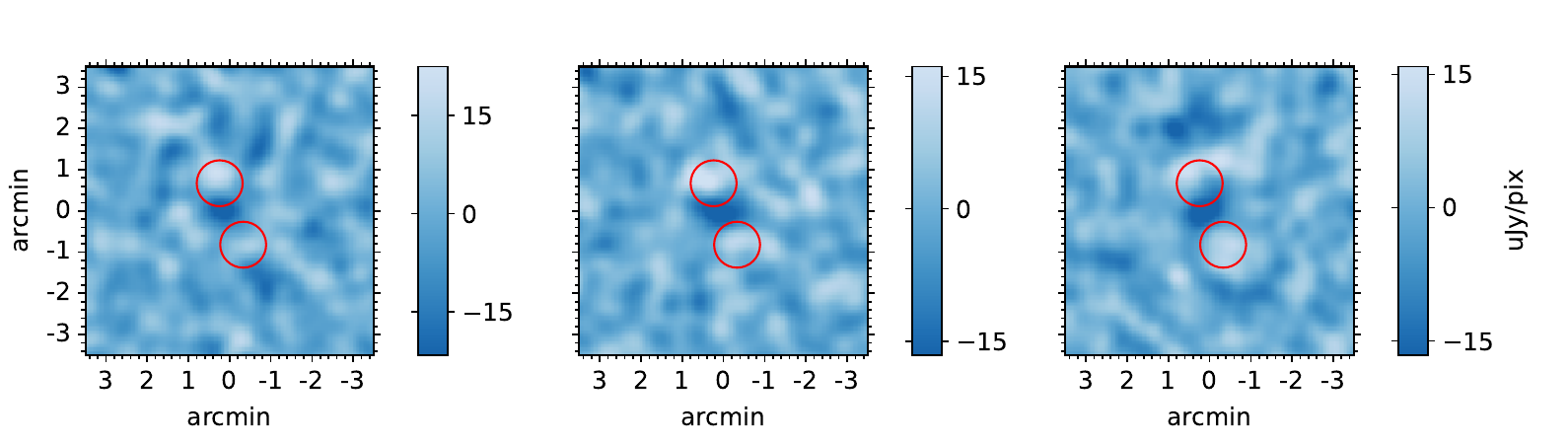}
     
   		\end{subfigure}
    	\begin{subfigure}
                \centering
        	\includegraphics[width=0.8\linewidth]{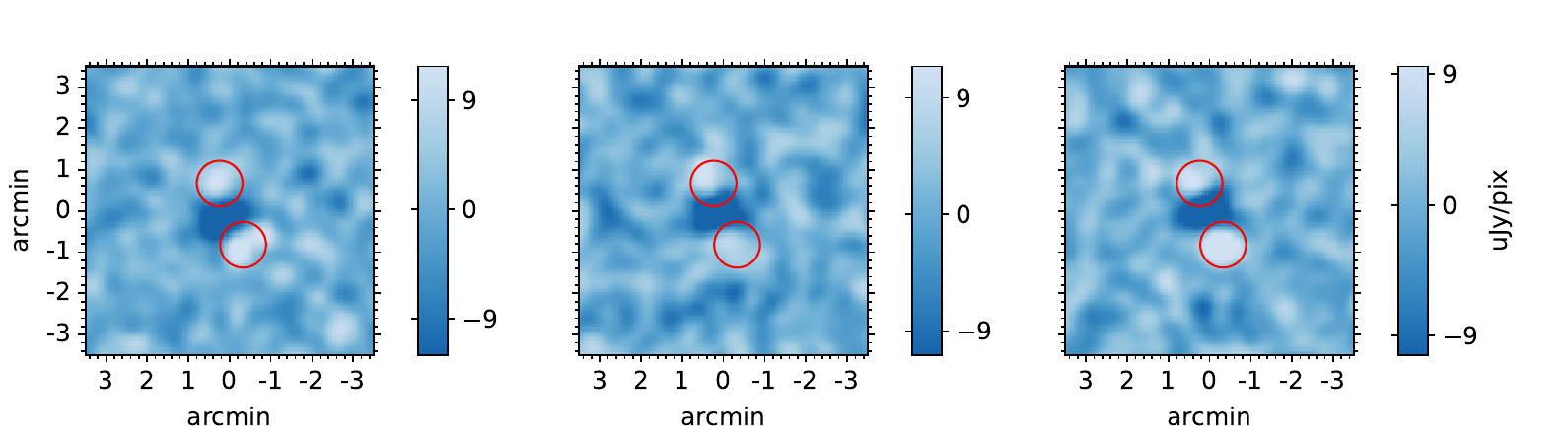}
        
   	 	\end{subfigure}
   	 	\begin{subfigure}
                \centering
        	\includegraphics[width=0.28\linewidth]{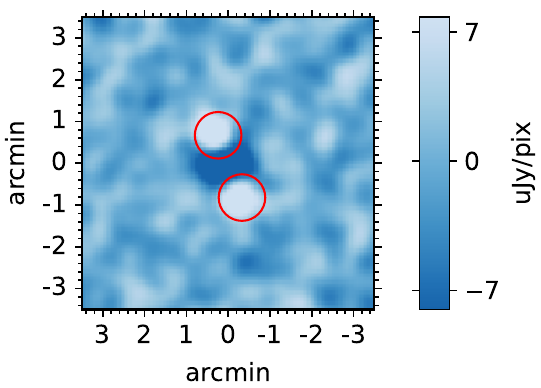}
        	
   	 	\end{subfigure}	
	\caption{CLEANed images of simulated SKA observations of the model cluster with non-thermal cavities, where $f_{14}=0.989$ and $p_\mathrm{1}=100$ (constrained by CARMA), representing an estimate of the sky brightness distribution seen by the SKA at $14.11$ GHz. Top panel (left to right): HA=0.5 days=1, HA=0.5 days=2, HA=1 days=1. Centre panel (left to right): HA=1 days=2, HA=2 days=1, HA=1 days=3. Bottom: HA=2 days=2. These images were CLEANed with CASA,  using a Gaussian taper of $3000\lambda$ to down-weight the longest baselines and $uvrange<10000 \lambda$ to reduce contributions from very small amplitude signal. A circular mask is also applied around the centre of the image with a radius of 30pix to isolate the cluster and the bubbles for the \emph{clean} algorithm. The red contours represent the true bubble positions and radius (as an offset).}
\label{fig:casa_nt}
\end{figure*}

Figure \ref{fig:casa_nt} shows the CLEANed images of 7 different observation lengths from \textsc{Profile} simulations (one of the 50 simulations that were produced). The observation times tested are 1, 2, 4, 6 and 8 hours over either 1, 2 or 3 days, defined by the hour angle (HA) and number of observing days. Note that an hour angle of 1 hour, for example, will mean that the telescope will begin observing one hour before the source crosses the meridian of the telescope, and will end one hour after it has crossed this point, giving a total of 2 hours observing time in one day. To make sure that the bubbles are extracted from the noise, a Gaussian taper weighting ($uv$taper parameter in CASA) is applied to the $uv$ data with a width of $3000\lambda$. This is used to reduce the weighting of the longest baselines so that the random fluctuations due to noise where the signal amplitude is very small are not as prominent. A circular mask is also applied around the centre of the image with a radius of $30$pix to enclose the cluster and the bubbles. 
The $uv$range is set to $< 10000 \lambda$ because the bubbles of cluster MS0735 are relatively large and the redshift is relatively low, the signal on the longest baseline lengths will be very faint. Figure \ref{fig:uvv} shows the $uv$-coverage of the SKA from these simulations.

The SZ signal from the cluster itself is mostly detected with each observation time, and the SZ contrast from the non-thermal bubbles becomes more clear the longer the SKA observes for. The CLEANed images show a first approximation via the image sky brightness that the SKA will be able to detect the MS0735 bubbles, but statistical analysis on the $uv$ data will indicate how accurate the suppression factor constraint is. Therefore, model comparison and parameter estimation are performed directly in $uv$ space. The data are prepared for Bayesian inference by binning the output visibilities from the simulation (with a $uv$ cut-off of $10000 \lambda$), with grid cells of size $130 \lambda$ for $14.11$ GHz observations. Bin sizes are determined based on the aperture illumination function (Fourier transform of the primary beam) size, following \cite{M}.

\begin{table*}
	\centering
	\caption{True values of cluster MS0735 parameters used in the model of the SZ signal map. The cavity positions ($\mathrm{x_{b}}$ and {$\mathrm{y_{b}}$)}} are given as offsets from the cluster center at (0,0). Also listed are the priors used in the Bayesian analysis. Cavity parameters are derived from \emph{Chandra} X-ray observations \citep{Vantyghem}. Subscripts $1$ and $2$ refer to the northern and southern bubble, respectively and $r_{b}$ is the bubble radius.
	\label{tab:params}
	\begin{tabular}{lcccr}
    	\hline
     	\headrow Parameter & True value (arcsec) & Prior type &  Prior value\\
    	$z$ & 0.216 & delta & 0.216\\
    	$\mathrm{x}_\mathrm{b, 1}$ (arcsec)  & 14.4 & uniform  & min=-50, max=50\\
    	$\mathrm{y}_\mathrm{b, 1}$ (arcsec) & 40.2 & uniform & min=0, max=100\\
    	$r_\mathrm{b, 1}$ (arcsec) & 30 & uniform & min=0, max=50\\
    	$\mathrm{x}_\mathrm{b, 2}$ (arcsec) & -19.8 & uniform  & min=-50, max=50\\
    	$\mathrm{y}_\mathrm{b, 2}$ (arcsec) & -49.2 & uniform & min=-100, max=0\\
    	$r_\mathrm{b, 2}$ (arcsec) & 30 & uniform & min=0, max=50\\
    	non-thermal $f_{14}$ & 0.989 & uniform & min=-2, max=2\\
    	$\theta_{s}$ (arcmin) & 5.3 & uniform & min=0.0833, max=10\\
    	$Y_\mathrm{tot}$ (arcmin$^{2}$) & 0.0051 & Gaussian & mean=0.0051, $\sigma$=0.001\\
    	\hline
  	\end{tabular}
\end{table*}

\begin{figure*}
  \centering
  \begin{subfigure}
  \centering
    \includegraphics[width=0.3\linewidth]{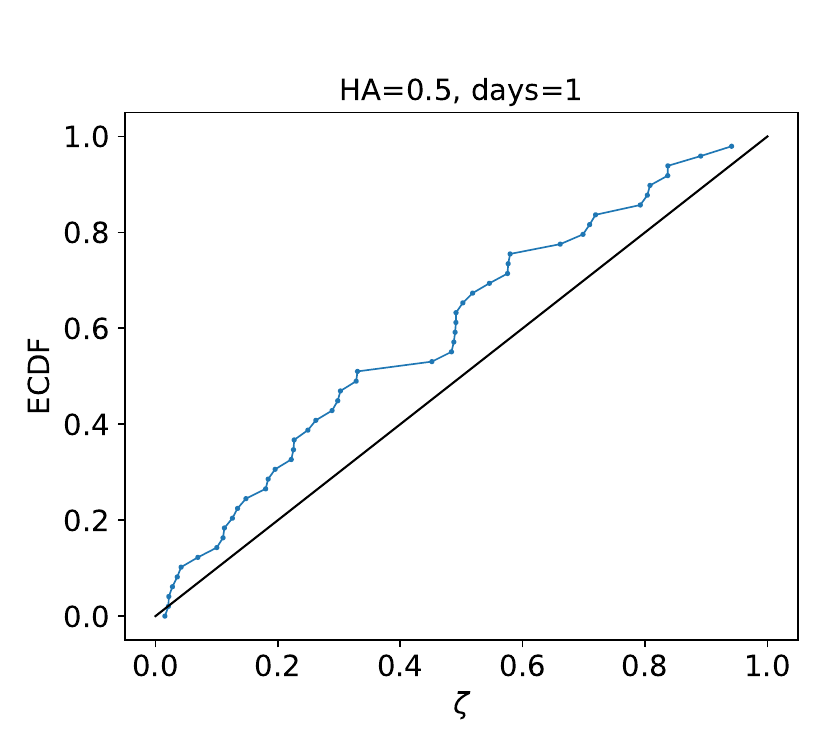}
  \end{subfigure}
  \begin{subfigure}
  \centering
    \includegraphics[width=0.3\linewidth]{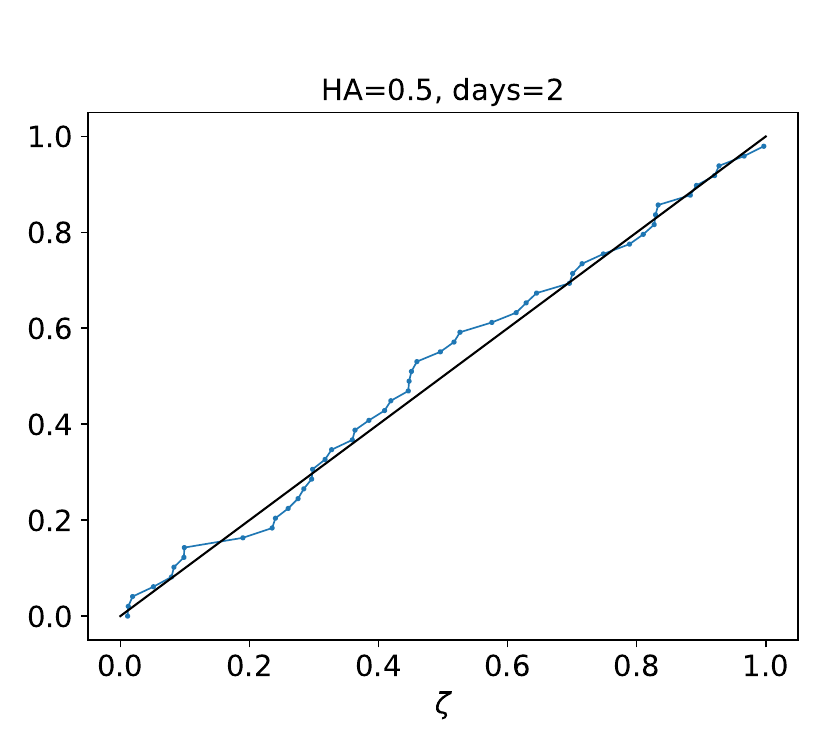}
  \end{subfigure}
  \begin{subfigure}
  \centering
    \includegraphics[width=0.3\linewidth]{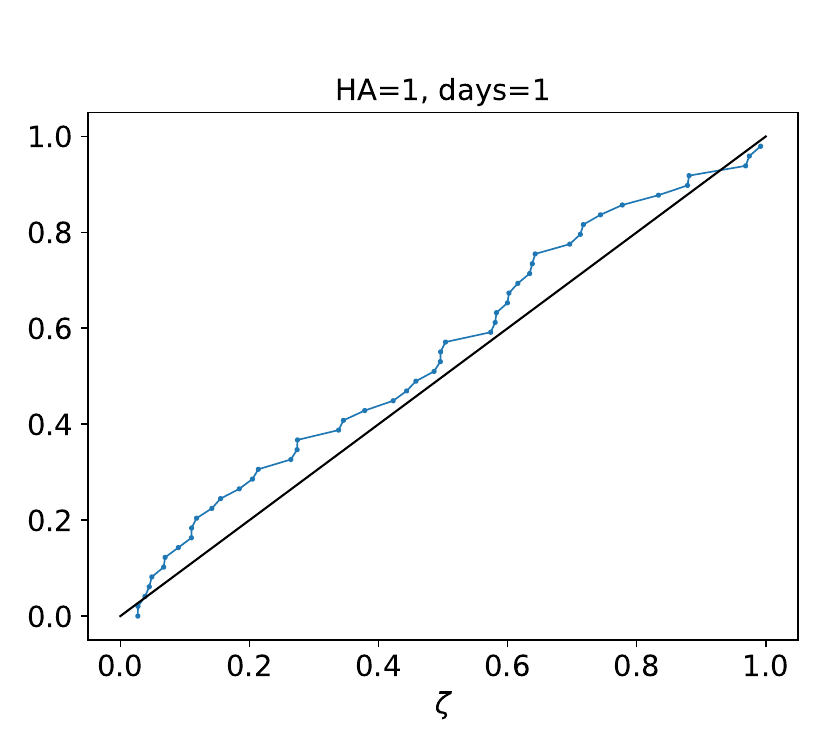}
  \end{subfigure}
  \begin{subfigure}
  \centering
    \includegraphics[width=0.3\linewidth]{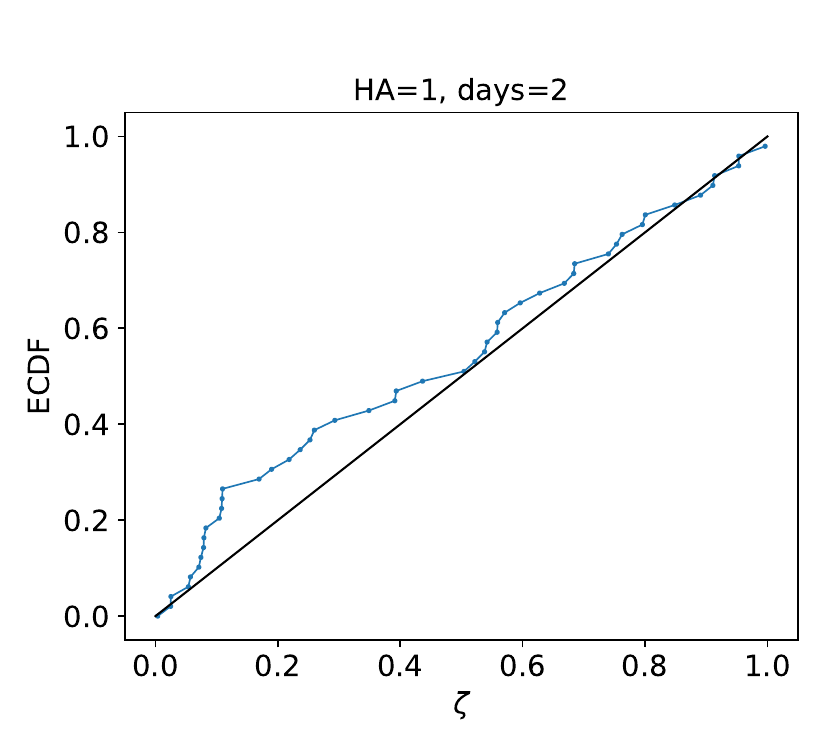}
  \end{subfigure}
  \begin{subfigure}
  \centering
    \includegraphics[width=0.3\linewidth]{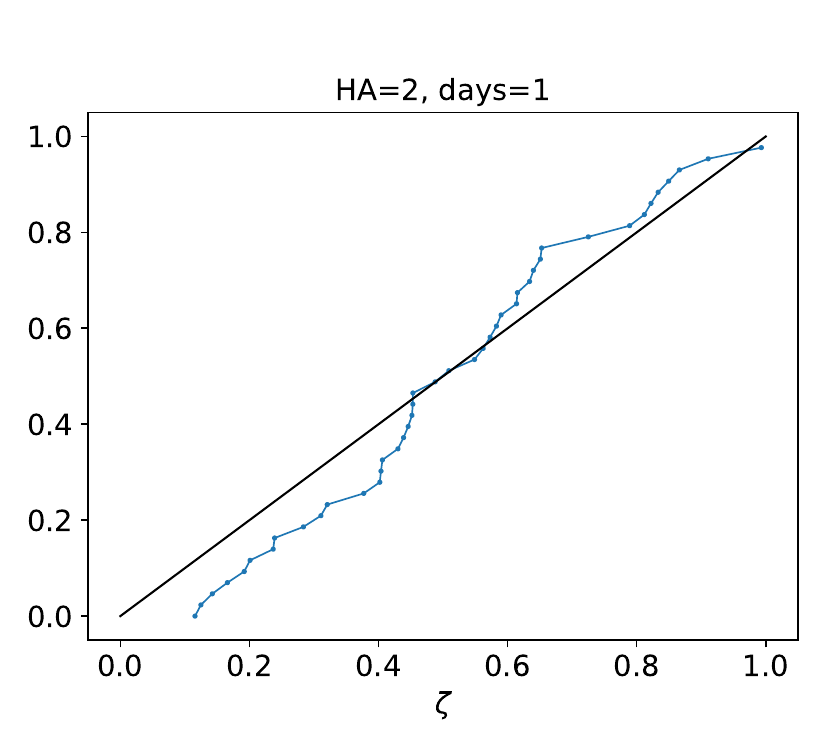}
  \end{subfigure}
  \begin{subfigure}
  \centering
    \includegraphics[width=0.3\linewidth]{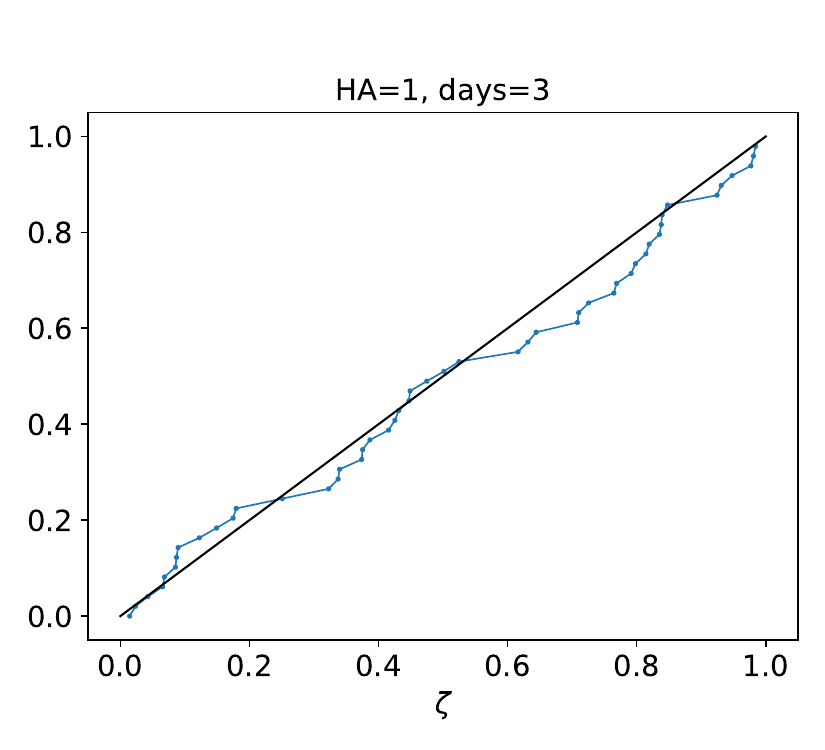}
  \end{subfigure}
  \begin{subfigure}
  \centering
    \includegraphics[width=0.3\linewidth]{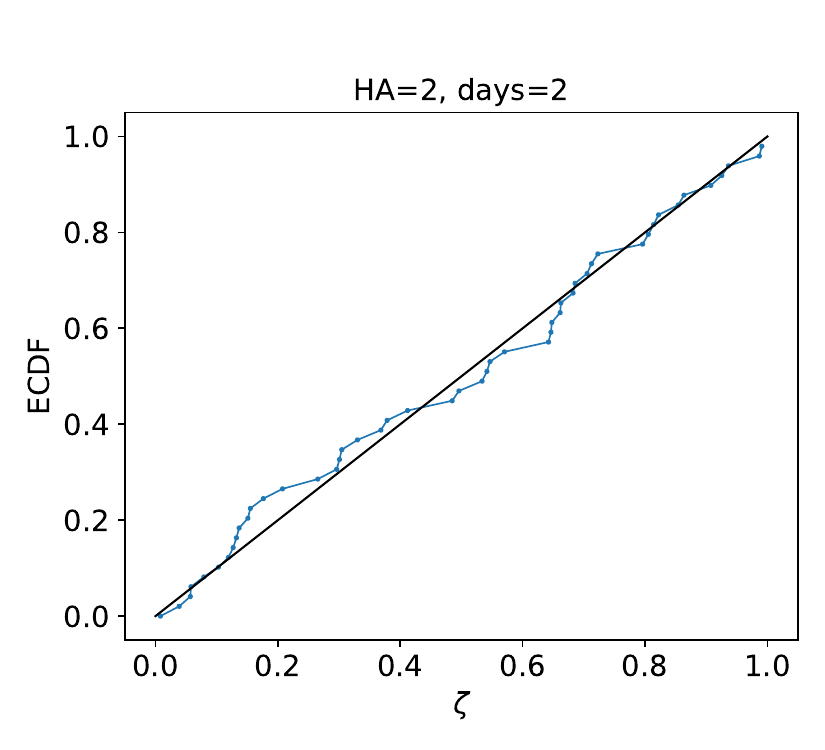}
  \end{subfigure}
  \caption{Suppression factor posterior validation curves of the non-thermal MS0735 cavities ($f_{14}=0.989$) produced from 50 simulations of each observation time. The error is overestimated for the 1 hour observation. The 2 hour curves fit well to the null hypothesis, therefore the error of the 50 posteriors is a good representation of the true constraint. Although the curves of the 4 hour observations are a worse fit to the null hypothesis, $\sim$68\% of the 50 posteriors contain the true value of $f$ in their 68\% confidence interval, which is expected from a good constraint. The curves for a 6 and 8 hour observation fit well, and the suppression factor has been well constrained.}
  \label{fig:nonthermalposval}
\end{figure*}

\begin{figure}
\centering
\includegraphics[width=\columnwidth]{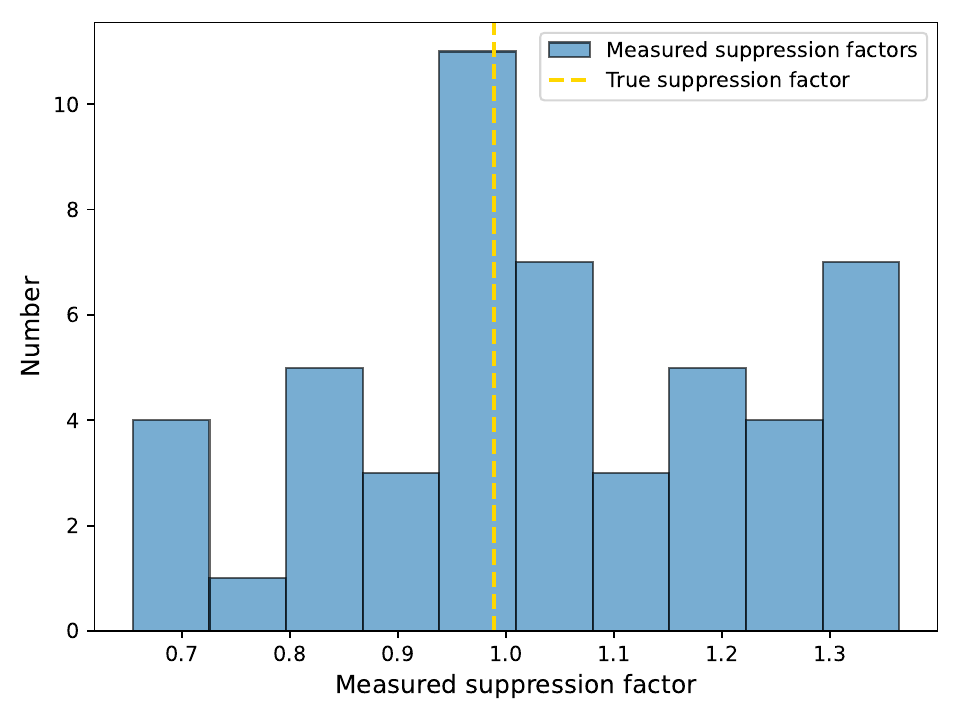}
\caption{The estimated suppression factors for the 50 realisations of simulated 8 hour observations of the model MS0735 non-thermal cavities. This shows that the estimates are scattered randomly around the input value.}
\label{fig:hist}
\end{figure}

The posterior validation curves resulting from the 50 suppression factor posteriors of each observation time are shown in Figure \ref{fig:nonthermalposval}. To have a realistic chance of observing the MS0735 cavities with the SKA in the future, an observation time of no longer than 6 or 8 hours should be able to constrain the suppression factor, as the demand for observing time is expected to be high. The posterior validation curve is a good fit to the null hypothesis curve/expected ECDF (black line) for both the 6 hour and 8 hour observations which implies that any deviations from the true value are consistent with the estimated error. Figure \ref{fig:hist} depicts a histogram of the estimated suppression factors for the 50 realisations of the 8 hour observation, alongside the true suppression factor value. This indicates that the estimates are scattered fairly randomly around the input. The top left image of Figure \ref{fig:nonthermalposval} suggests that an observation time of 1 hour is overestimating the error, therefore 1 hour can be ruled out as a viable observation time. The posterior validation curves of both 2 hour observations (top centre and right images in Figure \ref{fig:nonthermalposval}) are a better fit to the null hypothesis than the 1 hour observation and therefore deviations from the true value are more consistent with the error, and the fit is even comparable to the 6 and 8 hour observations. Therefore, a 2 hour observation would likely produce accurate measurements. Note that while the HA=0.5, days=2 case seems to be a better fit to the posterior validation curve than some of the longer observation times, we assume that this is likely due to a random fluctuation in the simulated noise. The posterior validation curve for the 4 hour observation over 2 days (left and centre images in the centre panel of Figure \ref{fig:nonthermalposval}) is a good fit to the null hypothesis. It is possible that the error is being underestimated with a 4 hour observation over 1 day (centre image in centre panel of Figure \ref{fig:nonthermalposval}), as the curve falls below the null hypothesis for low values of $\zeta$. However, $\sim$68\% of the posteriors contain the true value of the suppression factor within their 68\% confidence interval, indicating that the true value is being detected as expected for an accurate measurement.

The estimated suppression factors for 50 realizations of simulated 8-hour observations of the model MS0735 non-thermal cavities show no bias when compared to the input suppression factor.

\begin{figure}
\centering
\includegraphics[width=\columnwidth]{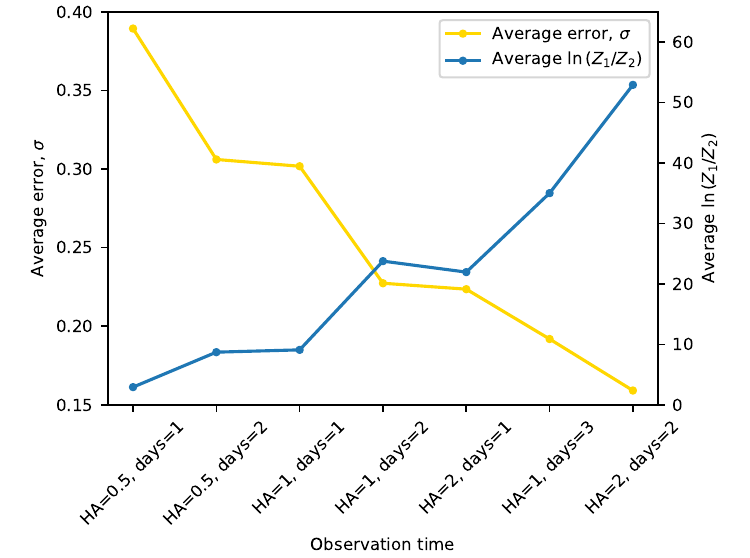}
\caption{The average error, $\sigma$, and $\ln(Z_{1}/Z_{2})$ of the 50 simulated SKA observations of non-thermal MS0735 cavities with $f_{14}=0.989$, for each tested observation time. The average $\ln(Z_{1}/Z_{2})$ increases with an increasing observation time, showing that the cavities become better detected. According to the scale in table \ref{tab:evidence}, there is very strong detection for each observation time, except for HA=0.5, days=1, where $\ln(Z_{1}/Z_{2})<5$. The error clearly decreases with an increasing observation time, showing the constraint on $f$ becomes much more informative. Note that while $\ln(Z_{1}/Z_{2})>5$ represents very strong detection of the cavities, it does not directly correspond to an accurate constraint on the suppression factor which instead requires $\ln(Z_{1}/Z_{2})>10$.}
\label{fig:evidencenonthermal}
\end{figure}

As described in section 2.4 the results from these observations are used to quantify a good constraint on the suppression factor. Figure \ref{fig:evidencenonthermal} depicts the average error $\sigma$ of the 50 suppression factor posteriors for each observation time, alongside the average values of $\ln (Z_{1}/Z_{2})$. Since one hour is ruled out as a viable observation time, this graph shows that $\ln(Z_{1}/Z_{2}) \approx 10$ is needed to represent a good suppression factor detection. The error that corresponds to this value is $ \sigma \approx 0.3$. The following results in this paper will use these thresholds to describe a good detection and an informative constraint on the suppression factor.

Next, we discuss the possibility of observing the suppression factor if it is thermal, and whether it could be distinguished from non-thermal suppression. Because of the similarity between the non-thermal and thermal suppression factors at $14.11$ GHz ($f_{14} = 0.989$ and $f_{14} = 0.977$ respectively), given the temperature and momentum values constrained by CARMA, it would be reasonable to assume that the above results also apply to the thermal scenario. However, \citealt{Abdullaetal} conclude that their suppression factor could define a thermal plasma of temperature `several hundred to thousands of keV'. Therefore, although their measured suppression factor corresponds to $\sim 10000$ keV, there could be a lower limit of $\sim 1000$ keV. It is possible that the temperature constrained by the suppression factor model is not exact, as there are many different features of the cluster that could shift the \emph{observed} suppression away from the \emph{true} suppression (which depends only on the distribution of electrons). For example, the ellipticity of the bubbles, the possibility of a shock affecting the SZ signal, mixtures of thermal and non-thermal electrons, and any differences in the pressure profile inside the bubble could all affect the \emph{observed} $f$ if they are not accounted for. It is then possible that $kT_{e}\approx1000$ keV and $p_{1}=100$ would give the same CARMA observed $f_{30}$ (even if they correspond to different true $f$ in the thermal and non-thermal models), if there are additional influences that are not accounted for. 

\subsection{Testing whether thermal and non-thermal SZ suppression can be distinguished}
Because of the above explanation, we next investigate whether a thermal and non-thermal suppression could be distinguished by the SKA if the CARMA constraint $kT_\mathrm{e}=1000$ keV is true, rather than $10000$ keV which gives an $f$ almost identical to the non-thermal case at $14.11$ GHz. If this is true, it is possible the plasma type could be discovered in future observations. 

\begin{figure}
\centering
\includegraphics[width=\columnwidth]{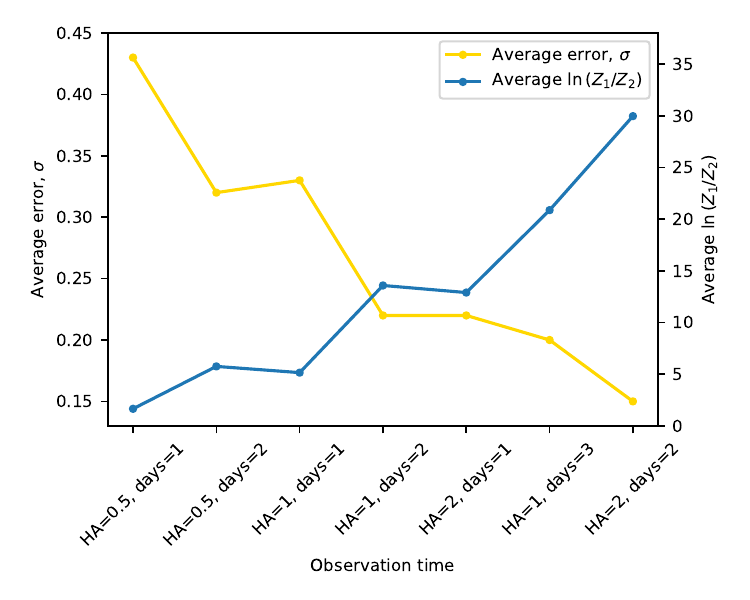}
\caption{The average error, $\sigma$, and $\ln(Z_{1}/Z_{2})$ of the 50 simulated SKA observations of thermal MS0735 cavities with $f_{14}=0.796$, for each tested observation time. The average $\ln(Z_{1}/Z_{2})$ increases with an increasing observation time, showing that the cavities become better detected. According to the scale in table \ref{tab:evidence}, there is very strong detection for each observation time, except for HA=0.5, days=1, where $\ln(Z_{1}/Z_{2})<5$. However, as described for the non-thermal case, this may not directly correspond to an accurate constraint on the suppression factor. The error clearly decreases with an increasing observation time, showing the constraint on $f$ becomes much more informative.}
\label{fig:evidencethermal}
\end{figure}

The suppression factor for $kT_\mathrm{e}=1000$ keV at $14.11$ GHz is $f_{14}=0.796$. A $y$-map of cluster MS0735 is modelled using this value. 50 \textsc{Profile} simulations for each of the 7 observation times are made, to mimic the SKA observations of the SZ signal. Figure \ref{fig:evidencethermal} shows the average value of $\ln(Z_{1}/Z_{2})$ and the average error, $\sigma$, of the 50 posteriors. The posterior validation curves for each observation time are not shown, as they are visually very similar to the non-thermal case, and the important features are quantified in Figure \ref{fig:evidencethermal}.

For both the 6 hour and 8 hour observations, the average $\ln(Z_{1}/Z_{2})>10$ and $\sigma<0.3$ which is the threshold for a good detection. We conclude from these results that a 6 and 8 hour observation time is sufficient to accurately constrain the suppression factor in the bubbles of galaxy cluster MS0735, in the thermal (lower limit of CARMA constrained $kT_\mathrm{e}$) case. Since the 1 hour observation time for the non-thermal case is not sufficient, the same is assumed for the thermal bubbles which have a lower suppression factor, and are therefore harder to detect. Note that although a smaller suppression factor means there is more signal coming from the bubbles, the SZ contrast between bubbles and global ICM is smaller, which is what makes the detection of $f$ more difficult. We conclude above that a 2 hour time could be viable for the non-thermal case. However, for the 2 hour thermal observation, $\ln(Z_{1}/Z_{2})<10$, which quantifies a poor constraint on $f$. An observation of thermal bubbles will need longer than 2 hours. For the more promising 4 hour observation, the average errors are both less than 0.30, which corresponds to the $\ln(Z_{1}/Z_{2})$ threshold. Therefore, a 4 hour observation of the thermal suppression factor in cluster MS0735 will give a well constrained value of $f$ for most observations.

To explore the SKA's ability to distinguish between a thermal and non-thermal scenario, given $kT_\mathrm{e}=1000$ keV or $p_{1}=100$, the average of mean values of the 50 posteriors for each observation time are compared in table \ref{tab:comp}. The 1, 2, 4 and 6 hour observation times have thermal and non-thermal average mean values within $1\sigma$ of each other. It is therefore unlikely that these observation times of the MS0735 cavities will result in distinguishable thermal and non-thermal $f$ values. The 8 hour observation time is able to produce measurements of the suppression factors that are just outside of $1\sigma$ from one another. While this shows that there is a slightly better chance of distinguishing between thermal and non-thermal suppression, a difference of $1\sigma$ still has a high probability of being due to random chance. It is therefore unlikely that the SKA will be able to distinguish between a thermal and non-thermal suppression factor in cluster MS0735 when observing for 8 hours or less.

\begin{table}
	\centering
	\caption{Average of the mean measured value of $f$ and its error $\sigma$ of the 50 posterior constraints for MS0735 bubbles with $p_{1}=100$ ($f_{14}=0.989$) in the non-thermal case, and $kT_\mathrm{e}=1000$ keV ($f_{14}=0.796$) in the thermal case. The 1, 2, 4 and 6 hour observations have thermal and non-thermal average mean values within $1\sigma$ of each other and cannot be distinguished. The 8 hour observation gives $f$ constraints that are just outside of $1\sigma$ from one another.}
	\label{tab:comp}
	\begin{tabular}{lccr}
    	\hline
     	\headrow & \shortstack{Total observation \\ time (hours)} & Non-thermal & Thermal \\
    	HA=0.5, days=1 & 1 & 1.09 $\pm$ 0.39 &  0.93 $\pm$ 0.43 \\
    	HA=0.5, days=2 & 2 & 1.08 $\pm$ 0.30  & 0.95 $\pm$ 0.32  \\
    	HA=1, days=1 & 2 & 1.10 $\pm$ 0.30  &  0.95 $\pm$ 0.33  \\
    	HA=1, days=2 & 4 & 1.07 $\pm$ 0.23  &  0.90 $\pm$ 0.22  \\
    	HA=2, days=1 & 4 & 1.04 $\pm$ 0.19  &  0.85 $\pm$  0.22  \\
    	HA=1, days=3 & 6 & 1.01 $\pm$ 0.22  &  0.87  $\pm$ 0.20  \\
    	HA=2, days=2 & 8 & 1.03 $\pm$ 0.16  &  0.81 $\pm$ 0.15  \\
    	\hline
  	\end{tabular}
\end{table}

To conclude, while the temperature and momentum constrained by CARMA from the suppression factor model ($p_{1}=100$ and $kT_\mathrm{e}=10000$ keV) do not yield distinguishable thermal and non-thermal $f_{14}$ values, the limited accuracy of the model allows for other possible values ($p_{1}=100$ and $kT_\mathrm{e}=1000$ keV) that correspond to suppression factors that the SKA has the potential to distinguish (albeit only at $1\sigma$ level) at 14.11 GHz.

\subsection{The Smallest Observable Suppression Factor}\label{sec:smallest}

As explained in section \ref{sec:intro} the constraints on the suppression factor measured by the GBT/MUSTANG-2 camera differ to those constrained by CARMA. The next investigation will explore how well the SKA can detect and distinguish thermal and non-thermal suppression factors of MS0735 cavities if they really are the values constrained by MUSTANG-2, and will in turn determine the smallest observable suppression factor in cluster MS0735.

\cite{Orlowski} constrain suppression factors ranging from $f_{90}=0.39$--$0.95$ and we create a simulated image of the cluster for five values of $f$ within this range: $f_{90}=0.39, 0.5, 0.6, 0.7, 0.8$. These values are used to constrain the possible values of $p_{1}$ given a non-thermal gas in the bubbles, and $kT_\mathrm{e}$ given a thermal gas. The following analysis is assuming a non-thermal gas. We obtain the corresponding suppression factors at the SKA frequency ($14.11$ GHz), $f_{14}=0.13, 0.29, 0.42, 0.57, 0.72$. Note that the range of suppression factors found by MUSTANG-2 correspond to $p_{1} \approx 1-15$ and $kT_\mathrm{e} \approx 100-3000$ keV. These temperature and momentum constraints differ to those found via CARMA observations, where $p_{1} \approx 100$ and $kT_\mathrm{e} \approx 10000$ keV. \citealt{Orlowski} investigated the impact of changing many assumptions, leading to a broad range of possible suppression factors compared to \citealt{Abdullaetal}. Therefore, the discrepancies are likely due to the simple model that ties the \emph{observed} suppression factor back to a $p_{1}$ or $kT_\mathrm{e}$ value, when it might not be the \emph{true} suppression factor. 

\begin{figure*}
		\begin{subfigure}
            \centering
        	\includegraphics[width=0.8\linewidth]{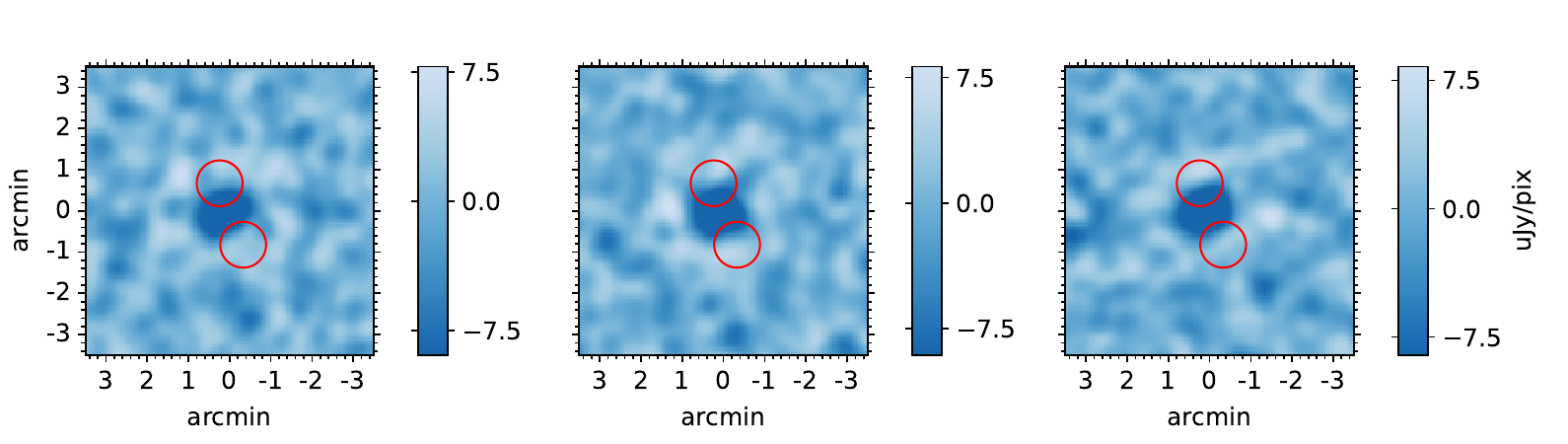}
     
   		\end{subfigure}
    	\begin{subfigure}
                \centering
        	\includegraphics[width=0.55\linewidth]{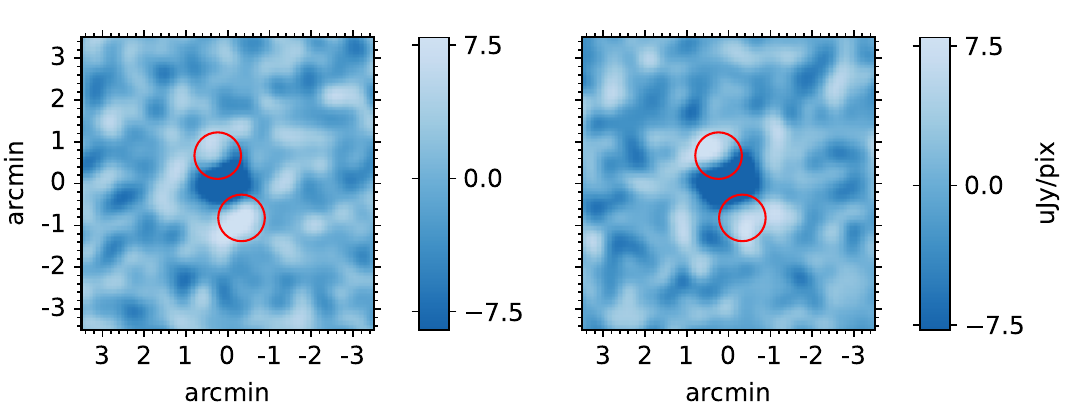}
        \end{subfigure}
	\caption{Images CLEANed via CASA showing simulated 8 hour SKA observations of cluster MS0735, with non-thermal cavities based on MUSTANG-2 constraints ($f_{90}=0.39, 0.5, 0.6, 0.7, 0.8$). These measurements constrain $p_{1}$ values, yielding $f$ values at the SKA frequency of $f_{14}=0.13$, $0.29$, $0.42$, $0.57$, $0.72$, depicted in these images. Top (left to right): $f_{14}=0.13$, $f_{14}=0.29$, $f_{14}=0.42$. Bottom (left to right): $f_{14}=0.57$, $f_{14}=0.72$. The images are produced using a Gaussian taper of $3000\lambda$ to down-weight the longest baselines, and a $uv$ cut-off of $10000\lambda$ to reduce contributes from very small amplitude signal. A circular mask is also applied around the center of the image with a radius of 30pix to isolate the cluster and the bubbles for the \emph{clean} algorithm. The red contours represent the true bubble positions and radius (as a offset).}
\label{fig:mustangcasa}
\end{figure*}

\textsc{Profile} simulations (CLEANed via CASA) in Figure \ref{fig:mustangcasa} show the SKA observed MS0735 non-thermal bubbles that are based on the minimum momentum constrained by the five MUSTANG-2 constrained suppression factors. This gives an initial indication that the bubbles are not observed for the lowest suppression factors. Each simulation is for an 8 hour observation only, so that the lowest possible detectable suppression factor can be investigated at the maximum viable observation time. 50 realisations of each observation are simulated, and Bayesian analysis is performed. 

\begin{figure}
\centering
\includegraphics[width=\columnwidth]{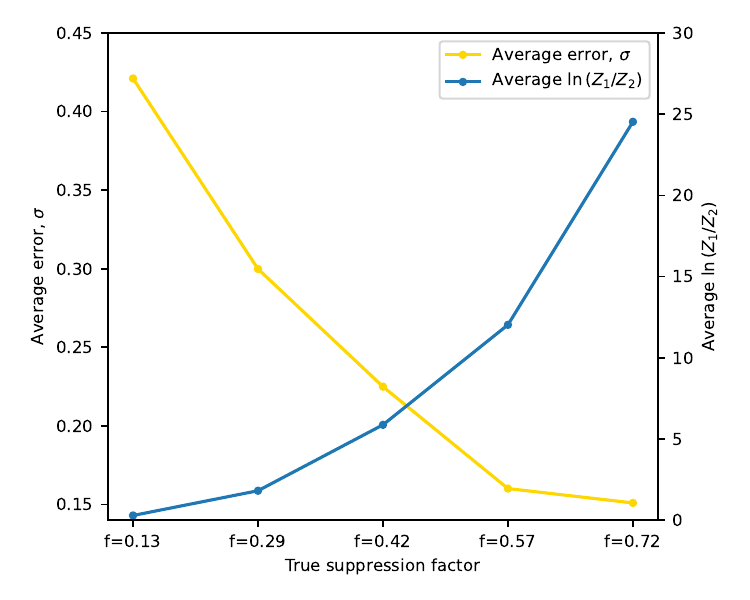}
\caption{The average error, $\sigma$, and $\ln(Z_{1}/Z_{2})$ of the 50 realisations of 8 hour observations of each possible value of non-thermal suppression factor ($f_{14}=0.13$, $0.29$, $0.42$, $0.57$, $0.72$) determined from MUSTANG-2 constrained $p_{1}$ values. The average $\ln(Z_{1}/Z_{2})$ increases with an increasing value of $f$, showing that the cavities become better detected with more suppression. According to the scale in table \ref{tab:evidence}, there is poor cavity detection when $f_{14}=0.13$ and $f_{14}=0.29$, as $\ln(Z_{1}/Z_{2}) < 5$, but cavities with the remaining $f_{14}$ values are strongly detected (note that a strong detection of the cavities does not directly imply an accurate measurement of the suppression factor; a value of $\ln(Z_{1}/Z_{2})>10$ needed). The error clearly decreases with an increasing value of $f_{14}$, showing the constraint on $f$ becomes much more informative if the SZ signal in the cavities is more suppressed.}
\label{fig:mustangevidence}
\end{figure}

The average value of $\ln(Z_{1}/Z_{2})$ and error, $\sigma$, on the $f$ constraint for the 50 realisations are given in Figure \ref{fig:mustangevidence}. Again, using previously mentioned thresholds, bubbles with $f_{14}=0.13$, $f_{14}=0.29$ and $f_{14}=0.42$ are not detected with an 8 hour observation time as the average $\ln(Z_{1}/Z_{2})<10$. Note that in this case, a value of $\ln(Z_{1}/Z_{2})>10$ corresponds to an error $<0.2$ rather than $<0.3$. This indicates that the error will need to be lower to represent a good constraint of a small suppression factor. This can also be understood by recognising that if the error is $0.3$ and the true value is $0.42$, then $f \approx 1.4 \sigma$ which is almost only one error bar away from $f=0$ (representing no bubble detection). So a smaller error is needed to pinpoint the true suppression. Observations of $f_{14}=0.57$ and $f_{14}=0.72$ both result in $\ln(Z_{1}/Z_{2})>10$ and $\sigma<$ 0.2.  Given these results, we conclude that the SKA can accurately measure the suppression factor of the MS0735 cavities if they are inhabited by non-thermal electrons with momentum that is constrained by MUSTANG-2, and that corresponds to suppression factors of $f_{14}=0.57$ and $f_{14}=0.72$. 

The five thermal suppression factor values at $14.11$ GHz that correspond to the range of cavity electron temperatures constrained by MUSTANG-2 are: $f_{14} = 0.27, 0.38, 0.49, 0.60, 0.73$. Using the results from the non-thermal case, the SKA would be able to detect $f_{14}=0.60, f_{14}=0.73$ and likely $f_{14}=0.49$, as these are larger than the corresponding non-thermal suppression factors that are well detected. Note that this conclusion is based on the results from the non-thermal simulations, and simulations with these exact suppression factor values were not performed. The thermal and non-thermal values at the SKA frequency are very similar, therefore we would not expect a differentiation between the two cases. This conclusion is supported by the following statistic: for the five non-thermal suppression factors observed by the SKA, the corresponding thermal suppression factor values fall within the 68\% confidence interval of at least 68\% of the 50 posteriors. Because of this, if the temperature and momentum within the cavities are indeed what was constrained from the MUSTANG-2 observations, the SKA would likely observe a value of $f$ that could represent both a thermal and a non-thermal gas. However, as described earlier, the suppression factor model ties the \emph{true} suppression factor to a $kT_\mathrm{e}$ and $p_{1}$, not the \emph{observed}. Therefore, different temperature and momentum values in the MS0735 bubbles are possible, and distinguishing between the thermal and non-thermal scenarios remains a possibility.

\subsection{Can the SKA measure the suppression factor of the MS0735 cavities?}

\vspace{5mm} In summary, the investigation of thermal and non-thermal suppression factors in the MS0735 cavities, considering existing constraints on electron temperature and momentum, has revealed that they will be observable by the SKA during an 8-hour observation, provided $f_{14} > 0.42$. If the cavities are described by the temperature $kT_\mathrm{e} \approx 10000$ keV, and momentum $p_{1} \approx 100$ constrained by CARMA, then an observation time as low as 4 hours will measure the suppression factor. It is also possible that a thermal and non-thermal suppression factor could be distinguished with an observation time greater than 8 hours, if instead $kT_{e} \approx 1000$ keV which is the lower end of the CARMA constrained temperature values. The range of $kT_\mathrm{e}$ and $p_{1}$ constrained by MUSTANG-2, however, imply that the suppression factors of the two gas types are too close to be distinguishable, if the (simple) suppression factor model is accurate.

\subsection{Cavity Line of Sight Position Effects}

An interesting effect on the suppression factor that was investigated in \cite{Orlowski} is the position of the bubbles along the LOS. It is likely that, in reality, the axis of the two AGN jets is not parallel to the plane of the sky, but is rotated and is inflating bubbles that are shifted along the LOS. The LOS geometry of the bubbles in cluster MS0735 is not known, therefore \citealt{Orlowski} investigate the influence on the suppression factor constraint, when increasing the angle $\theta$ from the plane of the sky. They find that the constrained value will increase with $\theta$. This is because the bubbles will move into more tenuous regions of the global ICM (given the radial dependence of pressure), therefore $y_\mathrm{cav}$ in equation \ref{eq:if} will decrease. The suppression factor $f$ must then increase to counteract the decrease in $y_\mathrm{cav}$, and to construct the signal received from MUSTANG-2. 

\vspace{5mm} This effect is also observed for constraints on the suppression factor by the SKA. We investigate this by exploring the effect of different LOS position priors on the MS0735 suppression factor. An 8 hour SKA observation is simulated to detect the signal of MS0735 constructed using $p_{1}=100$ in the non-thermal case. The bubbles are placed in the plane of the sky in the model image ($\theta=0$), and the true suppression factor is $f_{14}=0.989$. Bayesian inference is run on the data observed by the SKA for five different cases. First, an informative prior that the bubbles are in the plane of the sky is given. Then for each subsequent run, delta priors on $z_\mathrm{b}$ (in equation \ref{eq:z}) are given: $z_\mathrm{b} = 0.04, 0.08, 0.14, 0.24, 0.53$ Mpc for the northern bubble and $z_\mathrm{b} = -0.05, -0.10, -0.17, -0.30,$ $-0.64$ Mpc for the southern bubble. These correspond to LOS angles of $\theta = 15, 30, 45, 60, 75$. Note that asymmetry is due to the asymmetric $\mathrm{x}_\mathrm{b}$ and $\mathrm{y}_\mathrm{b}$ positions of the bubbles in the cluster. This analysis shows what would happen to the suppression factor constraint if we had an observed signal but did not know where the bubbles were along the LOS, so specified them to be at different (potentially incorrect) positions.

\begin{figure}
\centering
\includegraphics[width=0.8\columnwidth]{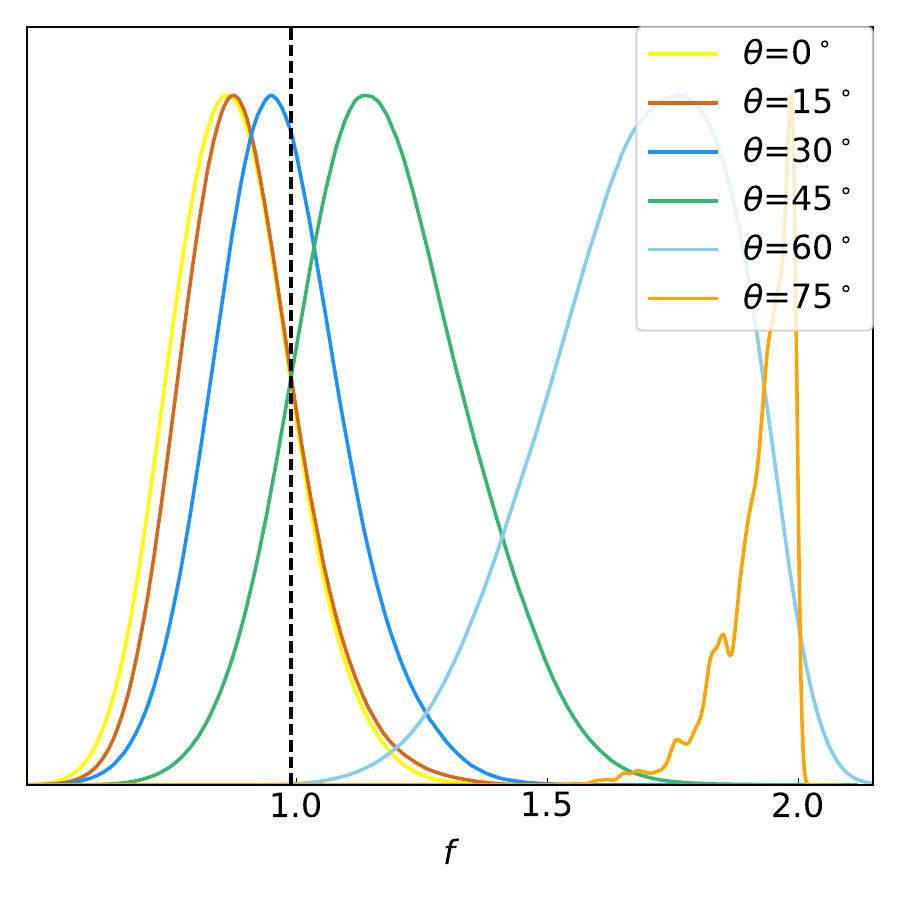}
  \caption{Posteriors of the constraints on non-thermal $f$. The simulated data were \emph{generated} with the bubbles at $z_\mathrm{b}=0$ (in the plane of the sky), but \emph{analysed} by assuming $z_\mathrm{b} = 0.04, 0.08, 0.14, 0.24, 0.53$ Mpc (northern bubble) and $z_\mathrm{b} = -0.05, -0.10, -0.17, -0.30,$ $-0.64$ Mpc (southern bubble) as delta priors. These shifts are related to angles $\theta = 15, 30, 45, 60, 75$. The dashed line represents the true value of $f_{14}=0.989$. This shows that if no prior information on the LOS position of the observed bubbles is known, it is possible they are in the plane of the sky and the measured $f$ will be smaller, or that they are along the LOS and the measured $f$ will be larger. Clearly, it is important to have prior information of the LOS position, so that the true $f$ can be discovered.}
\label{fig:losposteriors}
\end{figure}

The posterior distributions of the suppression factor for each of these LOS positions are shown in Figure \ref{fig:losposteriors}, where the true suppression factor value is represented by the vertical dashed lines. There is a cut off of the prior range at 2, which is causing the constraints of the largest line-of-sight angles to be restricted, with a less accurate shape. Similarly to the conclusions made regarding the LOS position for MUSTANG-2 observations in \cite{Orlowski} (see Figure 2 in their study), the SKA will measure larger suppression factors for a prior fixed at a larger value of LOS angle. This emphasizes the importance of knowing where the bubbles are along the LOS when analysing the observed signal. Incorrect assumptions of their position will lead to incorrect suppression factor measurements. We discuss this further in section \ref{sec:future} and suggest future investigations to help disentangle the degeneracy between suppression factor and LOS position. 

An important aspect of this analysis is that, for the SKA frequency ($14.11$ GHz), a value of $f_{14}>1$ in the cavities of MS0735 is not physical (e.g. the blue curves in Figure \ref{fig:f}). The original simulated $y$-map had a suppression factor of $f_{14}=0.989$ and thus the LOS angle shifts the measured suppression factor above $1$. This indicates that if the signal obtained from the MS0735 cavities via the SKA implies a suppression factor of $f_{14}=0.989$, under the assumption that the bubbles are located in the plane of the sky, then it would be unlikely that the bubbles are actually shifted some angle along the LOS. In the analysis of real cluster signals, priors that are based on the suppression factor values obtained from plots of $f$ against $kT_\mathrm{e}$ and $p_{1}$, should be used. This will put limits on the possible LOS positions of the bubbles for that given signal.

\begin{figure}
\centering
\includegraphics[width=\columnwidth]{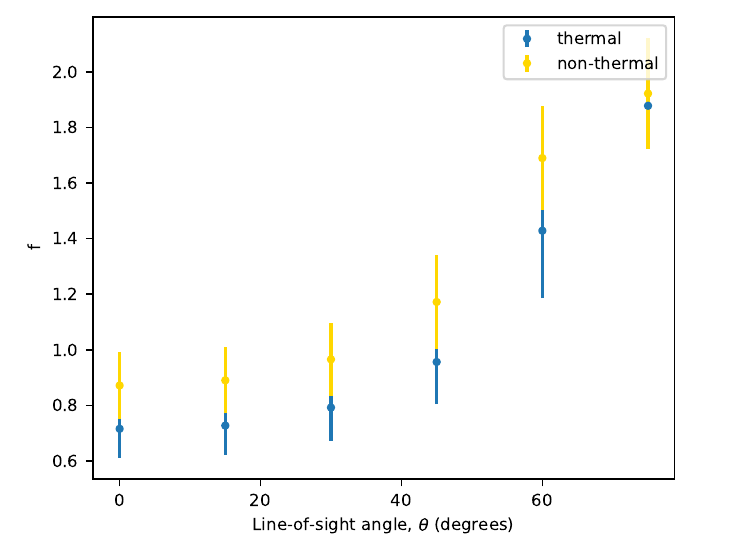}
\caption{Suppression factor, $f$, as a function of the line-of-sight angle $\theta$, with $\theta=0$ being in the plane of the sky and $\theta=90$ lying along the $z$-axis. The yellow data points represent measurement of the suppression factor given a non-thermal cavity gas with $p_{1}=100$ (as constrained for MS0735 by CARMA in \citealt{Abdullaetal}). The blue data points represent cavities with a thermal gas with $kT_\mathrm{e}=1000$ keV (the lower end of CARMA constraints). This shows that there is some degeneracy between thermal/non-thermal gas and LOS position, enforcing the importance of having prior LOS information so that the cavity gas type can be more accurately constrained.}
\label{fig:los}
\end{figure}

\vspace{5mm}Since the line-of-sight geometry is an added dimension to the model of the suppression factor within cluster cavities, it is important to consider the implications for being able to distinguish between the thermal and non-thermal scenarios. Figure \ref{fig:los} depicts the mean value and error of the suppression factor for the different LOS angles, measured with the Bayesian approach in the thermal and non-thermal case for MS0735 ($kT_\mathrm{e}=1000$ keV and $p_{1}=100$, respectively). While for each angle, there is still some difference between the thermal and non-thermal suppression factors (although with most of the errors overlapping), and the thermal values stay below the non-thermal ones, there will be degeneracy between the two scenarios for the smaller angles. For example, if the SKA were to measure a suppression factor of $\sim 0.9$, it is unclear whether this suppression factor represents thermal gas inhabiting bubbles which are at an angle of 45 degrees along the line-of-sight, or a non-thermal gas at an angle of 30 degrees. It is important for future observations to recognise this as an added challenge in determining the gas type. \cite{Orlowski} also note that this investigation reinforces the need for multi-wavelength SZ observations to disentangle the effects of different pressure support scenarios from the effect of LOS geometries. 

\begin{figure*}
		\begin{subfigure}
            \centering
        	\includegraphics[width=0.8\linewidth]{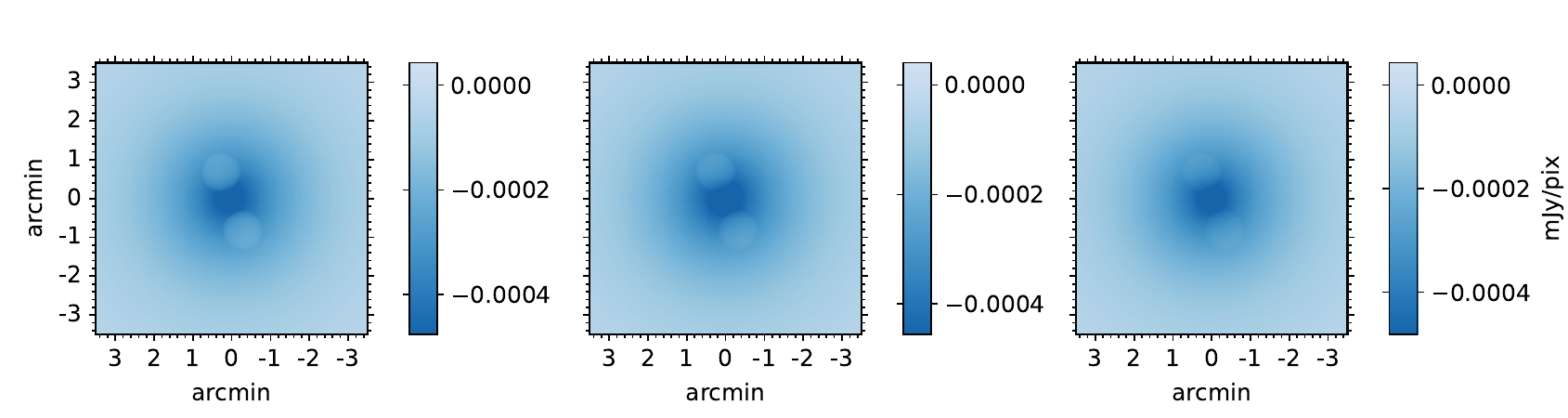}
     
   		\end{subfigure}
    	\begin{subfigure}
            \centering
        	\includegraphics[width=0.55\linewidth]{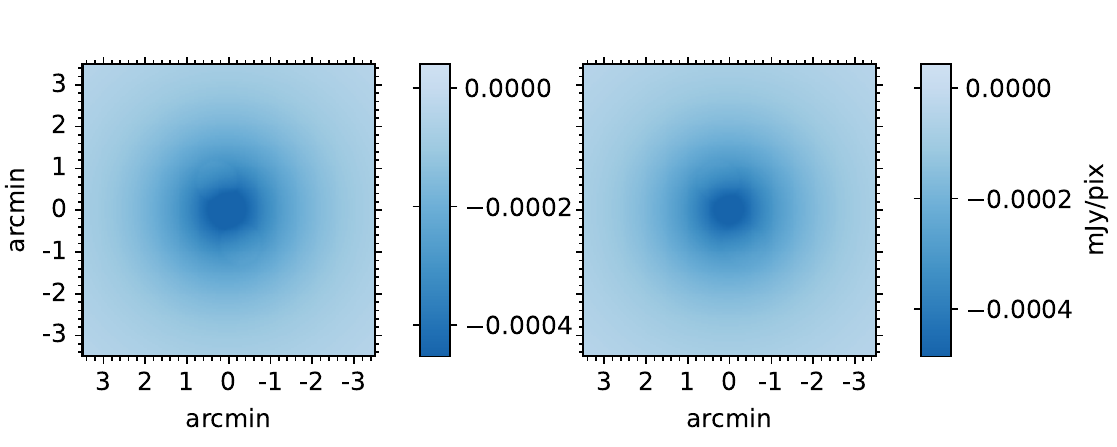}
        \end{subfigure}
	\caption{Model signal maps depicting non-thermal MS0735 cavities, at different angles along the LOS (the CARMA constraint of $p_{1}=100$ is used in each image and therefore $f_{14}=0.989$). Top panel (left to right): $\theta = 0$, $\theta = 45$, $\theta = 60$. Bottom panel (left to right): $\theta = 70$, $\theta = 75$. This could be misinterpreted as the suppression factor decreasing. In reality, the decrease in SZ contrast is because the LOS integration will cover more of the strong SZ signal at the cluster core, with the bubbles suppressing a weaker ICM region when shifted along the LOS. Therefore, longer observation times will be required to obtain an accurate measurement of $f$, if the bubbles are not in the plane of the sky.}
\label{fig:mapslos}
\end{figure*}

\vspace{5mm} Although the measured value of the suppression factor increases if the delta prior values on the LOS angle are higher (for the same observed signal), we discovered that detection of cavities and suppression factor constraints becomes much more difficult, the further along the LOS the cavities are positioned. Figure \ref{fig:mapslos}  depicts simulated SZ signal of cluster MS0735 based on non-thermal CARMA constraints in the cavities ($p_{1}=100$) for LOS angles $0, 45, 60, 70, 75$. It is clear the SZ contrast between cavity and ICM seems to decrease, despite the suppression factor staying the same in each image. This could be misinterpreted as the suppression factor decreasing, when in reality, it is because the LOS integration will cover more of the cluster core with strong SZ signal, so the signal received is enhanced. The bubbles will instead be suppressing a weaker ICM region, and will be less visible. This will affect the observation time required by the SKA to constrain accurately the true suppression factor. We performed an 8 hour SKA simulated observation of the non-thermal MS0735 cavities in the plane of the sky and at a LOS angle of $70$ degrees, as well as a 40 hour (HA=4, days=5) observation of the bubbles at $70$ degrees to test this theory. The average $\ln(Z_{1}/Z_{2})$ for the observations are: $\ln(Z_{1}/Z_{2})=$ 56.75 for an 8 hour observation of bubbles in the plane of the sky, $\ln(Z_{1}/Z_{2})=$ 1.55 for an 8 hour observation of bubbles at $\theta=70$, $\ln(Z_{1}/Z_{2})=$ 17.01 for a 40 hour observation of bubbles at $\theta=70$. It is clear that detection deteriorates if the bubbles are moved along the LOS, but improves with a longer observation time. A poor constraint via the SKA in the future could cause confusion regarding whether the suppression factor is too small to be detected, or whether it is large but hard to detect because the bubbles are out of the plane of the sky.

\section{Observing at Multiple Frequencies}\label{sec:freq}

Investigations in the preceding sections of this paper have included limitations such as a sole focus on the cavities of MS0735 which are the largest and most energetic cluster cavities known. This could limit the relevance of the previous results to other galaxy clusters, as most cluster cavities are a lot smaller. Additionally, clusters with cavities exist across a wide redshift and mass range. Therefore, it is imperative to this study that the suppression factor of a more diverse collection of cluster cavities is simulated and observed. This will increase the likelihood of observing the suppression factor in future observations and determining the nature of the cavity electrons. 

We also consider varying the observation frequency $\nu$. In the Rayleigh-Jean regime, the SZ spectrum has a $\nu^{2}$ dependence (with a negative gradient as the SZ effect in this region is represented by a deficit). This means the strength of the SZ signal that an instrument detects increases with the frequency of the observation. However, the optimum signal detection requires a balance between the increasing strength and decreasing antenna sensitivity with frequency. Additionally, the increase of frequency and corresponding decrease in wavelength $\lambda$ of the detected radiation results in the lengthening of baselines measured in $\lambda$. This will increase the sensitivity of the instrument to the signal of smaller scale structures at higher spatial frequencies. The interplay of these effects will be investigated below.

The investigation in this section involves calculating the number of 8 hour observing days required to constrain the non-thermal suppression factor, which is based on the MS0735 cavities with the CARMA constrained value of $p_{1} = 100$. We explore cavities ranging from $40$ to $80$ kpc, and each cluster has a mass of $M_{500}=5 \times 10^{14}M_{\odot}$. The positions of the bubbles are based off the MS0735 bubble positions. The radius of the MS0735 bubbles is approximately $100$ kpc. Therefore, when modelling clusters with bubble radii ranging from $40$ kpc to $80$ kpc, their positions are adjusted proportionally based on the percentage difference between their radii and the radius of the MS0735 bubbles. The true positions are given in table \ref{tab:freqparams}. 

\begin{table*}
\centering
\begin{tabular}{|>{\centering\arraybackslash}p{2cm}|c|c|c|c|c|c|}
   \hline
    \headrow Parameter (arcsec) & \multicolumn{3}{c|}{$z=1.5$} & \multicolumn{3}{c|}{$z=0.2$} \\
    \cline{2-7}
    \headrow & $r_{b}=40$ kpc & $r_{b}=60$ kpc & $r_{b}=80$ kpc & $r_{b}=40$ kpc  & $r_{b}=60$ kpc & $r_{b}=80$ kpc \\
    \hline
    $\mathrm{x}_\mathrm{b, 1}$ & 2.4 & 3.6 & 4.8 & 6.0 & 9.0 & 12.0\\
    $\mathrm{y}_\mathrm{b, 1}$ & 6.6 & 10.0 & 13.2 & 17.0 & 25.6 & 34.4 \\
    $r_\mathrm{b, 1}$ & 4.8 & 7.8 & 9.5 & 12.0 & 18.0 & 24.0 \\
    $\mathrm{x}_\mathrm{b, 2}$ & -3.3 & -4.9 & -6.5 & -8.4 & -12.0 & -16.8 \\
    $\mathrm{y}_\mathrm{b, 2}$ & -8.1 & -12 & -16.2 & -20.4 & -31.2 & -41.8 \\
    $r_\mathrm{b, 2}$ & 4.8 & 7.8 & 9.5 & 12.0 & 18.0 & 24.0 \\
    \hline
\end{tabular}
\caption{The positional parameters of the cavities used in this investigation. The coordinates of their centres on the model map of the sky, $\mathrm{x_{b}}$ and $\mathrm{y_{b}}$ (offset from the cluster center at (0,0)), are based on the MS0735 bubbles, which have a radius $\sim 100$ kpc. The positions of bubbles with radii $40$ kpc, $60$ kpc, $80$ kpc are adjusted proportionally based on the percentage difference between their radius and the radius of the MS0735 bubbles. Subscripts $1$ and $2$ refer to the northern and southern cavity respectively.}
\label{tab:freqparams}
\end{table*}

Three possible frequencies of the SKA are used to observe the clusters at redshift $z=0.2$ and $z=1.5$. These include the central frequency $23.75$ GHz of band $5+$ ($22.5-25$ GHz) and the central frequency $37.5$ GHz of band 6 ($36.25-38.75$ GHz) (these bands are both proposed), as well as $14.11$ GHz which has been used in the previous investigations of this paper. These frequencies are close together and therefore the suppression factor corresponding to $p_{1}=100$ is approximately the same for each, $f \approx 0.99$. At a redshift of $z=0.2$ and $z=1.5$, the corresponding scales are $3.30$ kpc/arcsec, and $8.46$ kpc/arcsec, respectively.

20 realisations of the observations are simulated. The FWHM of the primary beams used in these higher frequency observations are given in table \ref{tab:skabands}. The new SKA sensitivity calculator does not go above band 5. However, \cite{ska1} gives a predicted sensitivity model for the telescope, accounting for varying atmospheric, sky and receiver temperatures as a function of frequency. This was used in \cite{band6} to predict Band 6 sensitivity. Therefore, to find the new noise level for the band 5+ and band 6 observations, we scaled the \cite{ska1} model estimate by the same factor required to match the band 5 model estimate with the newly released sensitivity calculator. The noise levels used for each frequency are given in table \ref{tab:noise}.

\begin{table}
	\centering
	\caption{The noise level used in the \textsc{Profile} simulations for each frequency, including the adjusted noise level when the MeerKAT antennas are included (see section \ref{sec:improv})}
	\label{tab:noise}
	\begin{tabular}{lcr}
    	\hline
     	\headrow Frequency (GHz) & Noise level (Jy) \\
    	14.11 & 0.0061 \\
    	23.75 & 0.0110 \\
    	37.5 & 0.0160 \\
        14.11 (including MeerKAT antennas) & 0.0075 \\
    	\hline
  	\end{tabular}
\end{table}

An important difference to the analysis of the bubbles at higher redshift and smaller radii than MS0735, is that a much higher $uv$ cut-off is defined when binning the output \textsc{Profile} visibilities. For the highest redshift, a cut-off of $100$k$\lambda$ is used, and for the lower redshift, the cut-off is $50$k$\lambda$. These are chosen to ensure that no signal from longer baselines is cut-off, because observing smaller bubbles as well as at higher redshift, with higher frequency, means that more signal will appear at much higher baselines. For the $23.75$ GHz observation, the size of the grid cells used for binning the data are $220 \lambda$, and for $37.5$ GHz they are $340 \lambda$.

In the Bayesian analysis, we used a delta prior on the redshift so that the value was fixed. The prior on $Y_\mathrm{tot}$ was kept a Gaussian, similar to the MS0735 analysis, but the mean was changed to the true $Y_\mathrm{tot}$ value of the specific cluster, plus a 20\% error. The $\theta_{s}$ and $f$ priors are kept the same as the MS0735 analysis in table \ref{tab:params}. For $z=0.2$, the priors of the positional bubble parameters are kept the same as for MS0735 (as redshift and therefore angular scale is similar to MS0735). However, for $z=1.5$, the priors are narrowed due to the much smaller angular scale. These priors are given in table \ref{tab:priorsnew}.

\begin{table}
	\centering
	\caption{The prior types and values of the positional parameters used for the analysis of cluster bubbles at $z=1.5$. These are more narrow than the analysis at $z\sim 0.2$ because the angular scale is a lot smaller. The priors remain uninformative.}
	\label{tab:priorsnew}
	\begin{tabular}{lccr}
    	\hline
     	\headrow Parameter & Prior type &  Prior value\\
    	$\mathrm{x}_\mathrm{b, 1}$ (arcsec) & uniform  & min=-20, max=20\\
    	$\mathrm{y}_\mathrm{b, 1}$ (arcsec) & uniform & min=0, max=30\\
    	$\mathrm{y}_\mathrm{b, 1}$ (arcsec) & uniform & min=0, max=30\\
    	$\mathrm{r}_\mathrm{b, 1}$ (arcsec) & uniform & min=0, max=20\\
        $\mathrm{x}_\mathrm{b, 2}$ (arcsec) & uniform & min=-20, max=20\\
    	$\mathrm{y}_\mathrm{b, 2}$ (arcsec) & uniform & min=-30, max=0\\
    	$\mathrm{r}_\mathrm{b, 2}$ (arcsec) & uniform & min=0, max=20\\
    	\hline
  	\end{tabular}
\end{table}

\begin{figure*}
\centering
\includegraphics[width=0.45\textwidth]{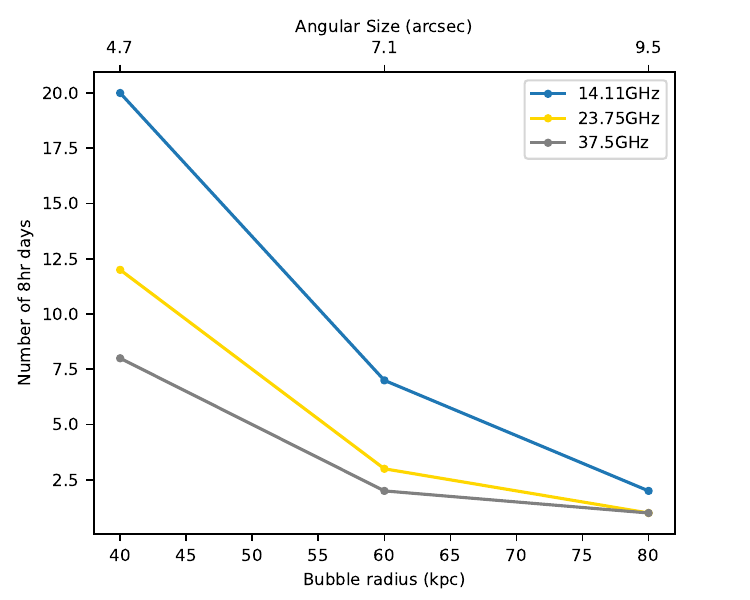}
\includegraphics[width=0.45\textwidth]{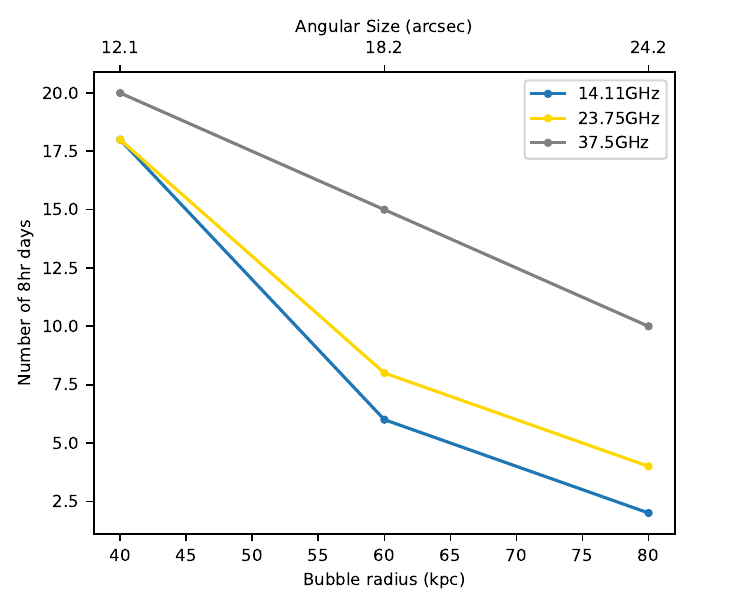}
\caption{Left: The number of 8 hour observing days required to detect the non-thermal suppression factor in a $z=1.5$ cluster for a range of bubble radii at three SKA frequencies. The highest frequency of $37.5$ GHz requires the shortest time for the smallest angular scale bubbles, due to its sensitivity to long baselines. However, for the larger angular scale bubbles, the increased noise at higher frequencies begins to balance the effect of the increased signal at long baselines. Then, each frequency requires approximately the same time to measure accurately the suppression factor for these bubbles. Right: The number of 8 hour observing days required to detect the non-thermal suppression factor in a $z=0.2$ cluster. For the smallest bubbles, each frequency requires a similar number of observing days. This makes sense because the angular size in arcsec is almost the same as the largest bubbles at $z=1.5$, where the three frequencies required similar times. The lowest frequency of $14.11$ GHz requires the shortest time for the largest bubbles, due to less signal occurring at higher baselines, and the smaller noise level that comes with lower frequency observations.}
\label{fig:z0p2}
\end{figure*}

The number of 8 hour observing days required to measure the suppression factor in different sized cavities for a cluster at $z=1.5$ and $z=0.2$ are shown in the left and right images of Figure  \ref{fig:z0p2}, respectively. The observation times are found by requiring that the average error of the 20 posteriors is $\lesssim 0.25$. This error is small enough to give an informative constraint on $f$. Another requirement is that the posterior validation curves lie approximately along the null hypothesis, so that the error is consistent with the deviation from the true value. 

The right image in Figure \ref{fig:z0p2} shows the number of 8 hour observing days required to constrain the suppression factor for bubbles with $40$ kpc, $60$ kpc and $80$ kpc radii, at redshift $z=0.2$. The trend for this lower redshift cluster shows that the three frequencies will require similar high observation times for bubbles with a $40$ kpc radius, and the curves will diverge as the cluster bubbles reach a radius of $80$ kpc. 

\vspace{5mm} The trend for $z=1.5$ (Left image in Figure \ref{fig:z0p2}) is that a higher observation frequency will require less time to acquire a good constraint on the suppression factor for the smallest bubble radii. Due to the lengthening of baselines at higher frequencies, more of the high spatial frequency $40$ kpc bubble signal is detected, resulting in a sufficient detection of the suppression factor at $37.5$ GHz in a shorter number of days than $23.75$ GHz or $14.11$ GHz. This shows that at high redshift with the smallest scale bubbles, the increased noise of higher frequency observations is compensated by the increased sensitivity to higher baselines and smaller structures. However, for the largest radius $r_\mathrm{b}=80$ kpc, each frequency requires a similar time. In this case, the higher frequencies don't improve the observation because a much smaller fraction of the total bubble signal lies at the higher baselines where the sensitivity is improved for higher frequencies, so the impact on the observation time is much less. Despite this, the redshift is high enough that there is still sufficient signal at higher spatial frequencies to compensate for the increased noise at the higher observation frequencies so that they require approximately the same observation time as $14.11$ GHz. The trend for the cluster at $z=0.2$, compared to the cluster at redshift $z=1.5$, makes sense by looking at the angular scale. The angular size in arcsec of the bubbles with the smallest radius at $z=0.2$ is almost the same as the largest bubbles at $z=1.5$, where the three frequencies also required similar times. The smallest angular scale of $4.7$ arcsec is best observed by the highest frequency, because of the sensitivity to small angular scales. The largest scale of $24.2$ arcsec is best observed by the lowest frequency due to the lack of signal at large baselines, and decrease in noise levels.

This investigation indicates that at redshifts $ \gtrsim 0.2 $, both frequency bands 5+ and 6, as well as the upper range of SKA band 5, could be suitable to detect small scale cluster bubbles ($r_\mathrm{b} \sim 40$ kpc). However, for the smallest angular scale bubbles in this investigation ($\sim 4.7$ - $9.5$ arcsec) a high frequency of $37.5$ GHz will produce the most efficient observation. Alternatively, for bubbles with angular scales $\gtrsim 12.1$ arcsec the SKA upper band 5 frequency of $\sim 14.11$ GHz will require a shorter and more suitable observation time. Note that most of the observation times for the high redshift and small scale bubbles are very long, requiring multiple 8-hour observing days. Observing the suppression factor of large angular scale cluster bubbles like cluster MS0735 is much more realistic when considering the allocation of SKA observing time. Despite this, the much shorter times required by the band 5+ and band 6 frequencies for high redshift and small scale bubbles provides a strong incentive for the expansion of SKA-Mid to higher frequencies, so that a more diverse range of cluster bubbles can be studied in the future.

\section{Improving SKA Observations}\label{sec:improv}

The previous investigations have excluded the MeerKAT antennas from the simulated observations, as it is not clear yet whether they will be fitted with band 5 receivers. This is identified as a possible limitation in the results presented so far. This section investigates the effect on the SKA's observation of the suppression factor in MS0735 cavities when the MeerKAT antennas are included, and acts to motivate the inclusion of the MeerKAT antennas for higher frequency observations in the future. 

For the purpose of this investigation, the effect of the improved $uv$-coverage is the main point of interest rather than sensitivity of the instrument, which is more difficult to implement in the simulations as there are three different types of baselines (SKA$\times$SKA, MeerKAT$\times$SKA and MeerKAT \\ $\times$MeerKAT). There is also an added complication that the MeerKAT antennas are not tested up to $14.11$ GHz, so the primary beam and noise levels would be only an extrapolation/guess if the three baseline types are included. Therefore, to implement the MeerKAT antennas, the primary beam is left fixed to the SKA primary beam (at 14.11 GHz, see table \ref{tab:skabands}). The sensitivity is assumed to be worse by a factor of $(13.5/15)^{2} = 0.81$, which is the ratio of the antenna collecting areas (see table \ref{tab:noise} for the noise level). It is then assumed that all of the baselines have this MeerKAT-MeerKAT sensitivity. This is a pessimistic estimate, as in reality the SKA-SKA and SKA-MeerKAT baselines will be more sensitive. Thus, if this estimation is helpful, we can be sure that the real observations will be the same or better. 

\begin{figure}
\centering
\includegraphics[width=0.9\columnwidth]{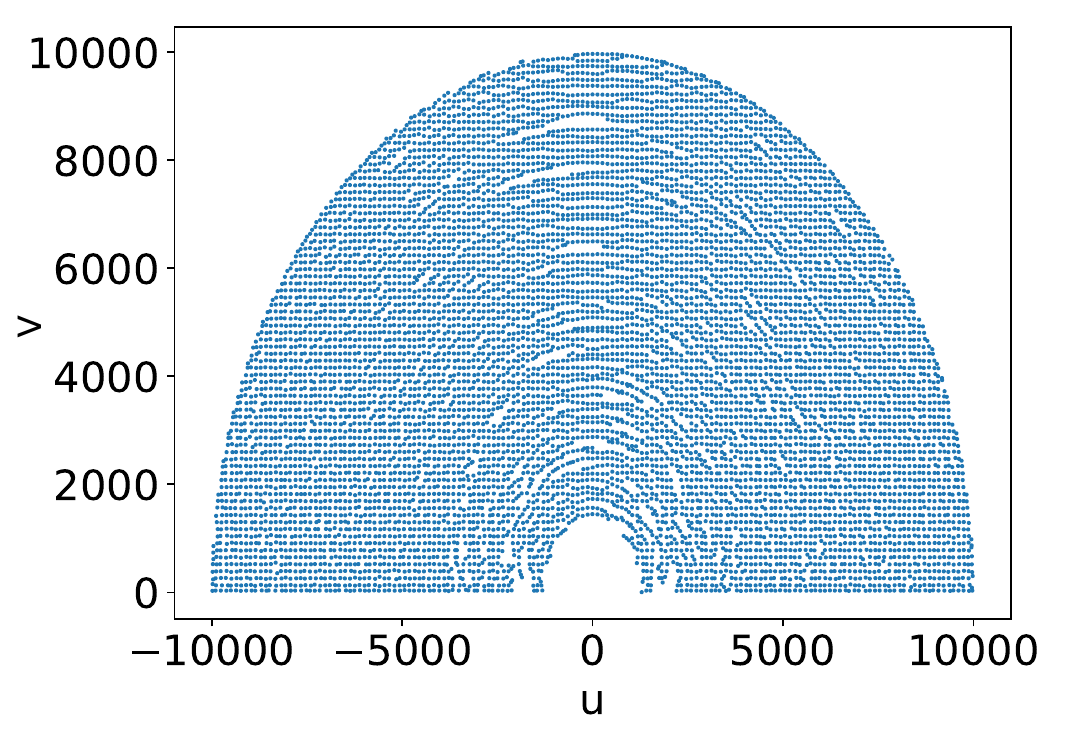}
\caption{The uv coverage in units of $\lambda$ (with a cut-off of 10000$\lambda$) of a simulated 8 hour SKA observation with MeerKAT antennas included.}
\label{fig:uvvv}
\end{figure}

\begin{figure}
\centering
\vspace{-10pt} 
\includegraphics[width=\columnwidth]{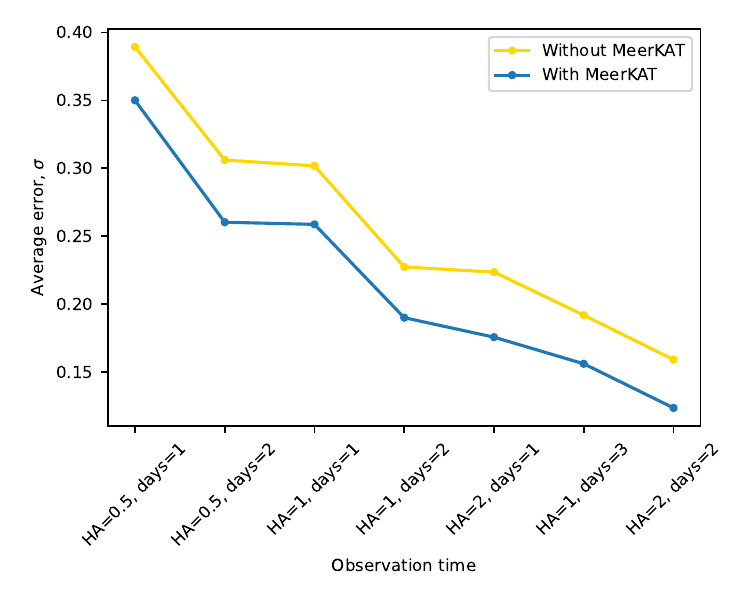}
\caption{Average error $\sigma$ of 50 posteriors for each observation time of simulated SKA observations of non-thermal MS0735 bubbles. The bubbles have $f_{14}=0.989$, based on the momentum constraint by CARMA, $p_{1}=100$. The yellow line is the same as the yellow curve in Figure \ref{fig:evidencenonthermal}, and the blue line is the result when the MeerKAT antennas are included in the observation.}
\label{fig:meerkaterror}
\end{figure}

\begin{figure}
\centering
\vspace{-10pt} 
\includegraphics[width=\columnwidth]{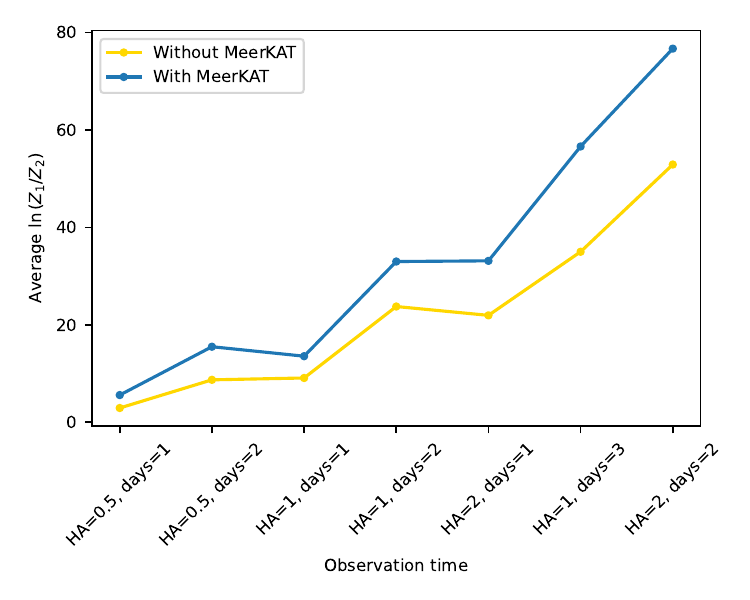}
\caption{The average $\ln (Z_{1}/Z_{2})$ of 50 realisations of each observation time of simulated SKA observations of non-thermal MS0735 bubbles. The bubbles have $f_{14}=0.989$, based on the momentum constraint by CARMA, $p_{1}=100$. The yellow line is the same as the blue curve in Figure \ref{fig:evidencenonthermal}, and the blue line is the result when the MeerKAT antennas are included in the observation.}
\label{fig:your_label}
\end{figure}

With this new simulation set up, we repeat the analysis from section \ref{sec:ms0735}, but now including the MeerKAT antennas. Figure \ref{fig:uvvv} shows the $uv$-coverage of the simulated SKA observations when the MeerKAT antennas are included. Figure \ref{fig:meerkaterror} shows the average values of $\sigma$ of the posteriors obtained from 50 realisations of each observation time, for MS0735 bubble observations. The yellow line represents observations without the MeerKAT antennas, and the blue line includes the MeerKAT antennas. Figure \ref{fig:your_label} shows the average of $\ln(Z_{1}/Z_{2})$, to represent the effectiveness of bubble detection. The error in the suppression factor constraint when the MeerKAT antennas are included is clearly smaller, and the value of $\ln(Z_{1}/Z_{2})$ is higher, implying that the bubbles of MS0735 are better detected. 

We have discussed throughout this paper that an average error below $0.3$ is associated with an observation time that can accurately constrain the suppression factor. The error for the 1 hour observation (HA=0.5, days=1) is still above this threshold. However, the error for both of the 2 hour observations (HA=0.5, days=2 and HA=1, days=1) is $\sim 0.26$, which is below the threshold, and is smaller than the error from the observation excluding MeerKAT. Additionally, $\ln(Z_{1}/Z_{2})>10$.  Therefore, it is likely that including the MeerKAT antennas for a 2 hour observation of the MS0735 suppression factor will produce a better constraint that is close to the true value, with a smaller error. The constraints on $f$ for the longer observation times are also improved. For example, if an error $\sigma<0.2$ is desired, an observation time of 4 hours (HA=1, days=2) would be possible if the MeerKAT antennas are included, but a 6 hour observation would be needed if they weren't. Similarly, to obtain an error $\sigma \approx 0.15$, including the MeerKAT antennas will require only a 6 hour observation, but without will require 8 hours. 

\begin{figure}
\centering
\vspace{-10pt} 
\includegraphics[width=\columnwidth]{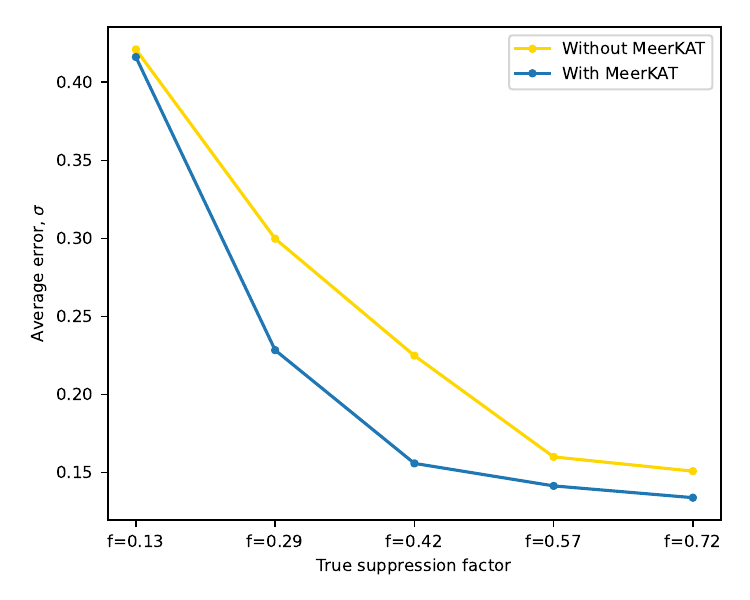}
\caption{Average error $\sigma$ of 50 posteriors from simulated SKA observations of a range of possible MS0735 suppression factors ($f_{14}=0.13$, $0.29$, $0.42$, $0.57$, $0.72$), that were derived from MUSTANG-2 constrained $p_{1}$ values. The yellow line is the same as the yellow curve in Figure \ref{fig:mustangevidence}. The blue line is the result when the MeerKAT antennas are included in the observation. The suppression described by $f_{14}=0.13$ is so small that including the MeerKAT antennas does not decrease the error on the constraint. The remaining suppression factors constraints are all improved by including MeerKAT antennas.}
\label{fig:lowesterror}
\end{figure}

As a final test of how well the MeerKAT antennas can improve a suppression factor detection, the simulations of section \ref{sec:smallest} are repeated. This determines if including the MeerKAT antennas will allow detections of smaller suppression factors in the future. Figure \ref{fig:lowesterror} shows the average error of 50 realisations of 8 hour observations of a range of suppression factors constrained by MUSTANG-2 (see section \ref{fig:lowesterror}). It is clear that including the MeerKAT antennas reduces the average error, especially for $f_{14}=0.29$ and $f_{14}=0.42$, improving the constraint on each value of $f$. However, for $f_{14}=0.13$ or $f_{14}=0.29$ the error is large relative to the true value, and it is clear that the SKA still cannot detect these suppression factors even when the MeerKAT antennas are included. 

We found in section \ref{sec:smallest} that the SKA likely will not detect $f_{14} = 0.42$. However, including the MeerKAT antennas has decreased the error on the constraint of $f_{14}=0.42$, to $<0.2$. This is promising, because as mentioned in section \ref{fig:lowesterror} the detection of a smaller suppression factor will require an error $\sigma<0.2$ to represent a good constraint. Additionally, the errors are $<0.2$ for the highest two suppression factors, $0.57$, $0.72$. Based on these results, if the MeerKAT antennas are included in the observation, it is likely that the SKA will be able to detect suppression factors $\geqslant 0.42$. 

\section{Limitations and Future Work}\label{sec:future}

Despite the promising predictions of the SKA's performance that have been presented in this work, further research is required to achieve the primary objective of constraining the contents of cluster cavities. 

First, since no calibration errors are assumed in the simulations in this work, this should be included in future work. We also assumed that the longer baselines of the SKA will be used to accurately subtract compact sources so that cluster and cavity detections are not impacted. The validity of this assumption should also be explored. The investigations should also be extended to more redshifts and masses. Only two cluster masses of $8\times 10^{14}M_{\odot}$ (MS0735) and $5\times 10^{14}M_{\odot}$, and three redshifts at $0.2$, $0.216$ and $1.5$ were investigated in this paper. In general, different  mass and redshift ranges will have to be investigated to quantify the required  observation time for each case.

\begin{figure}
\centering
\includegraphics[width=\columnwidth]{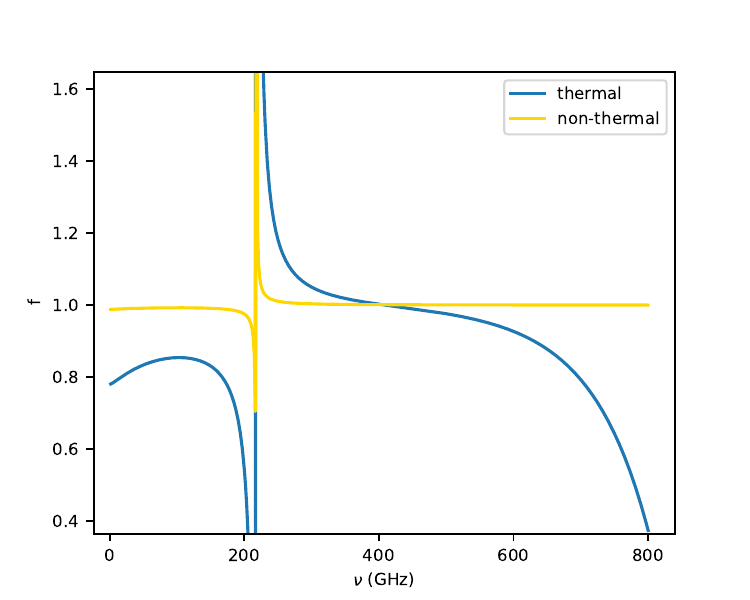}
\caption{Suppression factor spectrum of a non-thermal gas with $p_{1}=100$ (yellow curve), and a thermal gas with $kT_\mathrm{e}=1000$ keV (blue curve). Useful frequencies for future observations are $\nu \approx 200$ GHz and $250 \leq \nu \leq 300$ GHz, as these are close to the discontinuity where the curves from each gas type become more distinguishable.}
\label{fig:spec}
\end{figure}

One limitation in this research is that the frequencies investigated have produced thermal and non-thermal suppression factors that are very close in value. Observations at multiple different frequencies in the future will allow the measured values of $f$ to be tied back to the thermal or non-thermal theoretical spectrum. These observations should be performed at frequencies where the thermal and non-thermal values of $f$ are clearly distinct. This could include $\sim 200$ GHz (see Figure \ref{fig:spec}), where the curves start to depart as they head towards the discontinuity (where $g(x)$ approaches 0 in equation \ref{eq:supp}). Observing between $250$ GHz and $300$ GHz will also be useful, because the thermal suppression factor is allowed to be $>1$, but the non-thermal must stay at 1. Note that the spectra in Figure \ref{fig:spec} are based on $kT_\mathrm{e}$ and $p_{1}$ from CARMA measurements of MS0735, therefore $f$ may be allowed above 1 for other $p_{1}$ values. 

The frequencies suggested above are outside of the range of SKA-Mid frequency bands (see table \ref{tab:skabands}). Therefore, other instruments will likely be needed for future observations. Although CARMA has been decommissioned, ALMA is capable of observing at these frequencies. There are also some promising new instruments that could perform these observations, and whose results may be combined with those of the SKA to obtain a spectral dependence. The Atacama Large Aperture Sub-millimeter Telescope (AtLAST;\citealt{atlast}) is a concept for a 50m single-dish telescope that will provide high sensitivity and fast mapping of the (sub-)millimeter sky in the 2030s. AtLAST is expected to revolutionise SZ observations. Alongside high frequency observations, AtLAST will provide high angular resolution and increased sensitivity in comparison to existing single dish telescopes \citep{atlastsz} and therefore has the potential to effectively observe small-scale cavities, making it an excellent candidate for future suppression factor study.  While this is promising, further research and simulations should be carried out to understand the optimal observation parameters of AtLAST for accurate suppression factor measurement.

A second limitation that was encountered is the degeneracy between suppression factor and cavity LOS position. The signal that is obtained from cluster cavities can either be explained by a lower suppression factor with cavities in the plane of the sky, or a higher suppression factor if the cavities are assumed to be shifted along the LOS. Additionally, if the cavities really are along the LOS, much higher observation times are required for an accurate suppression factor measurement. If the LOS position of the observed cavities is known, the degeneracy can be broken, and an observation of suitable length can be made. 

\cite{kineticsz} suggest a possible solution. They find via simulations of cosmic ray (non-thermal) jet injection and subsequent bubble inflation, that inclined jets will induce a kinetic SZ signal which will be opposite in each cavity due to opposite LOS velocity components from the electrons. If the jets are ejected parallel to the plane of the sky, the signal is much weaker. Therefore, observing opposite signs of this signal in the two cavities will indicate whether they are in the plane of the sky, or shifted out. If the signal strength of the kinetic SZ effect can be directly linked to the specific angle of inclination through further research, this can be used as a prior in the Bayesian analysis of suppression factor measurement. The degeneracy will be broken and constraints will be much more accurate. To achieve this, the kinetic SZ signal will need to be disentangled from the global ICM and bubble ICM thermal SZ signal. Whether this is possible could be investigated in future work.

Finally, we recognise that while our assumption of a GNFW model for the global ICM pressure distribution is generally a reliable representation of the cluster shape, not all clusters that the SKA will observe may follow this distribution. For example, \cite{Abdullaetal} assume a  triaxial ellipsoidal $\beta$-model for the global ICM of cluster MS0735. In the case of a real observation, it is important to accurately model the global ICM and shape of the cavities, as an incorrect model could effect the estimation of the suppression factor by causing a decreased SZ contrast if the model bubble regions partially contain the global ICM. Therefore, future investigations should determine the impact of incorrect models on the estimation of the suppression factor from real observations.

\section{Conclusions}\label{sec:end}

This paper has aimed to provide insight into SKA's ability to accurately constrain the suppression factor of cluster cavities in future observations, adding to the spectral investigation of $f$, and aiding in the search for the cavity gas type. 

The most important discovery from the research presented in this paper is the potential for the SKA to provide an accurate constraint on the suppression factor of cluster MS0735 in as little as 4 hours. Additionally, given the constraints of $kT_{e}$ and $p_{1}$ from CARMA, a measurement of $f$ that differentiates between a thermal and non-thermal gas could be possible with an observation time greater than 8 hours. The SKA is one of the most promising radio astronomy projects, meaning limitations on observing times will be strict. 4 hours is likely short enough to justify the allocation of observing time. Because of this, a tighter constraint on the true value of $f$ in MS0735 is almost certain to come with SKA observations in the future, whose measurements can be compared to those that exist from CARMA and MUSTANG-2. 

The SKA will have another advantage, as a suppression factor as low as $f\approx0.42$ will be observable. This is important for future cluster observations, where cavities are likely less energetic than the well known MS0735 cavities, resulting in a less distorted CMB spectrum. The SKA's sensitivity to small values of $f$ is great motivation to observe new clusters, which might give a deeper insight into the cavity gas type. 

The proposed higher SKA frequencies of $37.5$ GHz and $23.75$ GHz will greatly enhance observations of small-scale clusters at $z \approx 1.5$. The observation times required to observe these types of cluster cavities will be much shorter than with $14.11$ GHz. These higher frequencies will not become available until a more advanced phase of the overall SKA project, but their effectiveness for cavity detection motivates their inclusion. 

The final results of this paper found that the MeerKAT antennas should be included in future observations. The increased $uv$-coverage of the array when they are included will decrease required observation times and allow a smaller suppression factor to be detected.

\begin{acknowledgement}
Computations were performed on the Rāpoi high-performance computing facility of Victoria University of Wellington.
\end{acknowledgement}

\paragraph{Funding Statement}

YCP is supported by a Rutherford Discovery Fellowship, managed by Royal Society Te Apārangi.

\paragraph{Competing Interests}
None.

\paragraph{Data Availability Statement}

Simulation and analysis code will be made available upon reasonable request to the author.

\printendnotes

\printbibliography

\end{document}